\documentclass[journal]{IEEEtran}
\usepackage[cmex10]{amsmath}
\usepackage{graphicx,amssymb,amsthm,comment,bm,cite,url}
\usepackage{color}
\usepackage{boldline}
\usepackage{algorithm,algorithmic}
\usepackage{multirow}
\usepackage{color}
\allowdisplaybreaks

\newcommand{\refeq}[1]{(\ref{eq:#1})}

\newcommand{\refsubsec}[1]{\ref{subsec:#1}}

\newcommand{\refsubsubsec}[1]{\ref{subsubsec:#1}}
\newcommand{\reffig}[1]{Fig. \ref{fig:#1}}

\newcommand{\reftab}[1]{Table \ref{tab:#1}}
\newcommand{\reftabs}[2]{Tables \ref{tab:#1} and \ref{tab:#2}}
\newcommand{\reftabss}[2]{Tables \ref{tab:#1}--\ref{tab:#2}}

\newcommand{\refalgo}[1]{Algorithm \ref{algo:#1}}
\newcommand{\refalgos}[2]{Algorithms \ref{algo:#1} and \ref{algo:#2}}

\def\Vec#1{\boldsymbol{\mathbf{#1}}}

\def\thline{\noalign{\hrule height 1.2pt}}

\ifCLASSINFOpdf
\else
\fi

\begin{document}

\title{
VoiceGrad: Non-Parallel Any-to-Many Voice Conversion with Annealed Langevin Dynamics
}

\author{Hirokazu~Kameoka,~
Takuhiro Kaneko,
Kou Tanaka,
Nobukatsu Hojo,
and Shogo Seki
\thanks{H. Kameoka, T. Kaneko, K. Tanaka, N. Hojo, and S. Seki 
are with NTT Communication Science Laboratories, Nippon Telegraph and Telephone Corporation, Atsugi, Kanagawa, 243-0198 Japan (e-mail: hirokazu.kameoka@ntt.com).}
}

%



\maketitle

\begin{abstract}
In this paper, we propose a non-parallel any-to-many voice conversion (VC) method termed {\it VoiceGrad}.
Inspired by WaveGrad, a recently introduced novel waveform generation method, VoiceGrad is based upon the concepts of score matching, Langevin dynamics, and diffusion models. 
The idea involves training a score approximator, a fully convolutional network with a U-Net structure, to predict the gradient of the log density of the speech feature sequences of multiple speakers. 
The trained score approximator can be used to perform VC by using annealed Langevin dynamics or reverse diffusion process to iteratively update an input feature sequence towards the nearest stationary point of the target distribution.
Thanks to the nature of this concept, VoiceGrad enables any-to-many VC, a VC scenario in which the speaker of input speech can be arbitrary, and allows for non-parallel training, which requires no parallel utterances.
\end{abstract}


%
\IEEEpeerreviewmaketitle

\section{Introduction}
\label{sec:intro}

Voice conversion (VC) is a technique to convert the voice of
a source speaker to another voice without changing the uttered sentence. 
Its applications range from
speaker-identity modification \cite{Kain1998short} to
speaking assistance \cite{Kain2007short,Nakamura2012short}, 
speech enhancement \cite{Inanoglu2009short,Turk2010short,Toda2012short}, 
bandwidth extension \cite{Jax2003short},
and accent conversion \cite{Felps2009short}.

While many conventional VC methods require 
parallel utterances to train acoustic models for feature mapping, 
{\it non-parallel} VC methods 
are ones that can work without parallel utterances or transcriptions for model training.
These methods can be useful in many cases
since constructing a parallel corpus is often very costly and non-scalable.
Another potentially important requirement for VC methods
is the ability to
achieve {\it any-to-many} conversion, namely to
convert speech of an arbitrary speaker to the voices of multiple speakers.
Such methods are also attractive in that they can work for input speech of unknown speakers without model retraining or adaptation.

A number of non-parallel methods have already been proposed, among which those that have attracted particular attention in recent years are based on deep generative models, 
such as variational autoencoders (VAEs) \cite{Kingma2014ashort,Kingma2014bshort}, generative adversarial networks (GANs) \cite{Goodfellow2014short}, and flow-based models \cite{Dinh2015,Dinh2017,Kingma2018}.

VAEs are a stochastic version of autoencoders (AEs), consisting of an encoder and decoder.
The encoder and decoder are modeled as different neural networks that produce a set of parameters
of parametric distributions, such as Gaussians.
The decoder represents the conditional distribution of a given set of data conditioned on a latent variable, whereas the encoder represents the posterior distribution of the latent variable.
In VAEs, the encoder and decoder networks are trained 
to maximize the variational lower bound of the marginal log-likelihood described by these distributions.
In VC methods based on VAEs \cite{Hsu2016short,Hsu2017short,vandenOord2017bshort,Huang2018,YSaito2018bshort,Kameoka2019IEEETransshort_ACVAE-VC},
the encoder is responsible for converting the acoustic features of input speech into latent variables, while the decoder is responsible for doing the opposite.
The basic idea is to condition the decoder on a target speaker code along with the latent variables so that the decoder can learn to generate acoustic features that are likely to be produced by the corresponding speaker and be linguistically consistent with the input speech.
Hence, these methods are capable of simultaneously learning mappings to multiple speakers' voices 
by using a single pair of encoder and decoder networks. 
By intentionally not conditioning the encoder on a source speaker code, the encoder can be trained to work in a speaker-independent manner. 
Under this setting, these methods allow for any-to-many conversions. 
Subsequent to these methods, a regular (non-variational) AE-based method called AutoVC was proposed \cite{Qian2019} and proved to be capable of handling any-to-any conversions by having the decoder take as input the speaker embeddings obtained with a speaker encoder pretrained using the generalized end-to-end loss \cite{Wan2018}.

GANs provide a general framework for training a generator network without an explicit definition of the generator distribution.
The goal is to train
a generator network so as to deceive a discriminator network, which learns to distinguish fake data generated by the generator from real data. 
The training process in GANs is formulated as a minimax game using an adversarial loss,
in such a way that the generator progressively gets better at generating data that appear to be real, while the discriminator gets better at distinguishing them as fake data. 
The minimax game using the adversarial loss 
is shown to be equivalent to a process of fitting the implicitly defined generator distribution to the data distribution.
We previously reported a non-parallel VC method \cite{Kaneko2018,Kaneko2019short_cycleganvc2} using a GAN variant called cycle-consistent 
GAN (CycleGAN) \cite{Zhu2017short,Kim2017short,Yi2017short}, which
was originally proposed as a method for translating images with unpaired training examples. 
The idea is to train a pair of mappings between one speaker's voice and another speaker's voice 
using a cycle-consistency loss along with the adversarial loss.
While the adversarial loss is used to ensure that the output of each mapping will follow the corresponding target distribution, 
the cycle-consistency loss is introduced to ensure that converting input speech into another speaker's voice and 
converting it back to the original speaker's voice will result in the original input speech.
This encourages each mapping to make only a minimal change from the input so as not to destroy the linguistic content of the input.
The cycle-consistency loss has recently proved 
effective also in VAE-based methods \cite{Tobing2019short}. 
Although the CycleGAN-based method was found to work reasonably well, 
one limitation is that it can only handle one-to-one conversions. 
To overcome this limitation, 
we further proposed an improved version \cite{Kameoka2018SLTshort_StarGAN-VC,Kaneko2019short_starganvc2,Kameoka2020IEEE-ASLP_StarGAN}
based on another GAN variant called StarGAN \cite{Choi2017short}, which 
offers the advantages of VAE-based and CycleGAN-based methods concurrently. 
As with VAE-based methods,
this method is capable of simultaneously learning mappings to multiple speakers' voices 
using a single network 
and thus can fully use available training data collected from multiple speakers. 
In addition, it works without source speaker information, and can thus handle any-to-many conversions.

Flow-based models are a class of generative models consisting of multiple invertible nonlinear layers 
called flows. 
Flows can be seen as a series of changes of variables, which
gradually transform each real data sample into a random noise sample following some prespecified distribution. 
The basic idea is to enable the direct evaluation and optimization of a log-likelihood function, which is usually hard to compute, 
by using a special network architecture consisting of flows whose Jacobians and inverse functions 
are easy to compute.
Recently, a non-parallel VC method using flow-based models has been proposed \cite{Serra2019}.
The principle is conceptually similar to VAE-based methods in the sense that 
the forward and inverse flows play similar roles as the encoder and decoder in a VAE.

Several VC methods based on sequence-to-sequence (S2S) models have also been  
proposed, including the ones we proposed previously \cite{Tanaka2019short,Kameoka2020_ConvS2S-VC,Huang2019arXiv_VTN,Kameoka2020_VTN}. 
While this approach usually requires parallel corpora for training, the recognition-synthesis approach \cite{Zheng2016,Sun2016,Miyoshi2017,Liu2018a,Liu2018b,Zhang2019,Zhang2020,Liu2021}, in which 
an automatic speech recognition (ASR) model and a decoder are cascaded to perform VC,
allows for nonparallel training by separately training the ASR model and decoder using text or phoneme transcripts. 
In this approach, the ASR model is trained to extract a sequence of linguistic-related features, e.g., phonetic posteriorgram (PPG) or a sequence of bottleneck features (BNFs) from source speech, whereas the decoder is trained to generate speech of a target speaker from that sequence.  
It should be noted that the top-performing systems in Voice Conversion Challenge (VCC) 2020 \cite{Zhao2020} adopted the PPG-based recognition-synthesis approach. 

Score-based generative models \cite{Song2019,Song2020} or diffusion probabilistic models \cite{Ho2020,Nichol2021} are another notable class of generative models, different from the above, that have recently been proven to be very effective in generating images and speech waveforms.
Inspired by the success of these models, 
in this paper we propose yet another method for non-parallel any-to-many VC based on the concepts of Langevin dynamics and reverse diffusion, and compare it objectively and subjectively with the conventional methods.
The proposed model uses a neural network in a way that the behavior depends less on the distribution of inputs,
which we expect to be advantageous in any-to-one or any-to-many VC tasks, especially under low-resource conditions.
Another motivation for adopting a score-based generative model or DPM for VC is the flexibility to customize the conversion process to meet various user requirements. We anticipate achieving this by combining independently pretrained classifiers or other types of score-based models to adjust the update direction at each time step of the Langevin dynamics or reverse diffusion process. This aspect is particularly appealing because it allows for customization without requiring retraining.

\section{Score Matching and Lengevin Dynamics}
\label{sec:scorematching}

We start by briefly reviewing the fundamentals of the score-based generative models, i.e., the concepts of Langevin dynamics and score matching. 

For any continuously differentiable probability density $p(\Vec{x})$, 
we call $\nabla_{\Vec{x}} \log p(\Vec{x})= \frac{\partial \log p(\Vec{x})}{\partial \Vec{x}}$ 
its {\it score function} \cite{Hyvarinen2005}.
If we are given the score function of the data of interest, 
we can use Langevin dynamics to draw samples from the corresponding distribution:
Starting from an initial point
$\Vec{x}$, 
we can 
iteratively refine it 
in a noisy gradient ascent fashion 
so that the log-density $\log p(\Vec{x})$ will be increased
\begin{align}
\Vec{x} \leftarrow \Vec{x} + \gamma \nabla_{\Vec{x}} \log p(\Vec{x}) +\sqrt{2\gamma} \Vec{z},
\label{eq:langevin}
\end{align}
where $\gamma > 0$ is a step size
and $\Vec{z}$ is a zero-mean Gaussian white noise with variance 1 that is drawn independently at each iteration. 
It can be shown that 
when $\gamma$ is sufficiently small and the number of iterations is sufficiently large, 
$\Vec{x}$ will be an exact sample from $p(\Vec{x})$ under some regularity conditions. 
This idea is particularly attractive in that we only need to access 
$\nabla_{\Vec{x}} \log p(\Vec{x})$ instead of $p(\Vec{x})$, which is usually very hard to estimate.
Hence, given a set of training examples $\mathcal{X} = \{\Vec{x}_n\}_{1\le n\le N}$, 
the focus is on how to estimate the score function from $\mathcal{X}$ at hand.

Score matching \cite{Hyvarinen2005} is a method to 
estimate the score function of the true distribution by optimizing 
a score approximator $\Vec{s}_{\theta}(\Vec{x})$ parameterized by $\theta$. 
We can use  
the expected squared error between $\Vec{s}_{\theta}(\Vec{x})$ and 
$\nabla_{\Vec{x}} \log p(\Vec{x})$
\begin{align}
\mathcal{E}(\theta) =
\mathbb{E}_{\Vec{x}\sim p(\Vec{x})}\left[
\left\|
\Vec{s}_{\theta}(\Vec{x}) - 
\nabla_{\Vec{x}} \log p(\Vec{x})
\right\|_2^2
\right],
\label{eq:score_matching}
\end{align}
as the objective function to be minimized with respect to $\theta$.
Here, $\mathbb{E}_{\Vec{x}\sim p(\Vec{x})}[\cdot]$ can be approximated 
as the sample mean over $\mathcal{X}$.
Even when 
the regression target $\nabla_{\Vec{x}} \log p(\Vec{x})$ is not directly accessible,
there are several ways to make this problem tractable 
without requiring an explicit expression of $p(\Vec{x})$.
One is implicit score matching \cite{Hyvarinen2005}, which uses 
the fact that \refeq{score_matching} is equivalent up to a constant to
\begin{align}
\mathcal{I}(\theta) =
\mathbb{E}_{\Vec{x}\sim p(\Vec{x})}\left[
2{\rm tr}(\nabla_{\Vec{x}}\Vec{s}_{\theta}(\Vec{x}))
+
\|
\Vec{s}_{\theta}(\Vec{x}) 
\|_2^2
\right],
\label{eq:implicit_score_matching}
\end{align}
where $\nabla_{\Vec{x}}\Vec{s}_{\theta}(\Vec{x})$ denotes the Jacobian of 
$\Vec{s}_{\theta}(\Vec{x})$, and 
${\rm tr}(\cdot)$ denotes the trace of a matrix.
However, unfortunately, computing
${\rm tr}(\nabla_{\Vec{x}}\Vec{s}_{\theta}(\Vec{x}))$
can be extremely expensive when $\Vec{s}_{\theta}(\Vec{x})$ is expressed 
as a deep neural network
and $\Vec{x}$ is high dimensional.
Another technique involves 
{\it denoising score matching (DSM)} \cite{Vincent2011}, which is noteworthy
in that it can completely circumvent ${\rm tr}(\nabla_{\Vec{x}}\Vec{s}_{\theta}(\Vec{x}))$.
The idea is to first perturb $\Vec{x}$ in accordance with a pre-specified noise distribution
$q_{\sigma}(\tilde{\Vec{x}}|\Vec{x})$ parameterized by $\sigma$ 
and then estimate the score of the distribution 
$q_{\sigma}(\tilde{\Vec{x}}) = \int q_{\sigma}(\tilde{\Vec{x}}|\Vec{x}) p(\Vec{x}) \mbox{d}\Vec{x}$
of the perturbed version.
It should be noted that $q_{\sigma}(\Vec{x})$ can be seen as 
a Parzen window density estimator for $p(\Vec{x})$. 
If we assume the noise distribution
to be Gaussian 
$q_{\sigma}(\tilde{\Vec{x}}|\Vec{x})=\mathcal{N}(\tilde{\Vec{x}};\Vec{x},\sigma^2\Vec{I})$,  
the loss function to be minimized becomes
\begin{align}
\mathcal{D}_{\sigma}(\theta) 
&=
\mathbb{E}_{\Vec{x}\sim p(\Vec{x}), \tilde{\Vec{x}}\sim \mathcal{N}(\tilde{\Vec{x}};\Vec{x},\sigma^2\Vec{I})}
\left[
\left\|
\Vec{s}_{\theta}(\tilde{\Vec{x}}) -
\frac{\Vec{x}-\tilde{\Vec{x}}}{\sigma^2}
\right\|_2^2
\right].
\label{eq:denoising_score_matching}
\end{align}
As shown in reference \cite{Vincent2011}, the optimal $\Vec{s}_{\theta}(\Vec{x})$
that minimizes \refeq{denoising_score_matching} almost surely equals 
$\nabla_{\Vec{x}}\log q_{\sigma}(\Vec{x})$, and it matches 
$\nabla_{\Vec{x}}\log p(\Vec{x})$ when 
the noise variance
$\sigma^2$ is small enough such that 
$ q_{\sigma}(\Vec{x})\approx p(\Vec{x})$.
The underlying intuition is that the gradient of the log density at some
perturbed point $\tilde{\Vec{x}}$ should be directed towards the original clean sample $\Vec{x}$.

Recently, attempts have been made to apply  
the DSM principle to image generation \cite{Song2019,Song2020}
by using a neural network to describe the score approximator $\Vec{s}_{\theta}(\Vec{x})$.
A major challenge to overcome 
in applying DSM to image generation tasks 
is that 
real-world data including images tend to reside on a low dimensional manifold 
embedded in a high-dimensional space.
This can be problematic in naive applications of DSM 
since 
the idea of DSM is valid only when the support of the data distribution is the whole space.
In practice, 
the scarcity of data in low-density regions can cause difficulties
in both score matching-based training and Langevin dynamics-based test sampling. 
To overcome this obstacle, 
the authors of \cite{Song2019}
proposed a DSM variant called {\it weighted DSM}.
The idea is to use 
multiple noise levels $\{\sigma_l\}_{l=1}^{L}$ in both training and test sampling.
During test sampling,
the noise level is gradually decreased 
so that $q_{\sigma_l}(\Vec{x})$
can initially fill the whole space and eventually converge to the 
true distribution $p(\Vec{x})$.
To let the score approximator learn to behave differently 
in accordance with the different noise levels, 
they proposed using 
a noise conditional network to describe $\Vec{s}_{\theta}(\Vec{x},l)$,
which takes the noise level index $l$ as an additional input.
For the training objective, they proposed using
a weighted sum of $\mathcal{D}_{\sigma_1}(\theta),\ldots,\mathcal{D}_{\sigma_L}(\theta)$
\begin{align}
\mathcal{L}_{\rm DSM}(\theta) = 
\frac{1}{L}
\sum_{l=1}^{L}\lambda_l \mathcal{D}_{\sigma_l}(\theta),
\end{align}  
where $\lambda_l>0$ is a positive constant that can be chosen arbitrarily.
Based on their observation that when $\Vec{s}_{\theta}(\Vec{x},l)$ is trained to optimality, 
$\|\Vec{s}_{\theta}(\Vec{x},l)\|_2$ tends to be proportional to $1/\sigma_l$,
they recommended setting $\lambda_l$ at $\sigma_l^2$, which results in
\begin{align}
\mathcal{L}_{\rm DSM}(\theta) =
\frac{1}{L}
\sum_{l=1}^{L}
\mathbb{E}_{\Vec{x},\tilde{\Vec{x}}}
\left[
\left\|
\sigma_l
\Vec{s}_{\theta}(\tilde{\Vec{x}},l) -
\frac{\Vec{x}-\tilde{\Vec{x}}}{\sigma_l}
\right\|_2^2
\right],
\label{eq:weighted_denoising_score_matching}
\end{align}
where the expectation is taken over 
the training examples of $\Vec{x}$ and the random samples of $\tilde{\Vec{x}}\sim \mathcal{N}(\tilde{\Vec{x}};\Vec{x},\sigma_l^2\Vec{I})$.
Note that they also recommended setting $\{\sigma_l\}_{1\le l\le L}$ to a positive geometric sequence 
such that $\frac{\sigma_2}{\sigma_1}=\cdots=\frac{\sigma_{L}}{\sigma_{L-1}}\in [0,1]$.
Once $\Vec{s}_{\theta}(\Vec{x},l)$ is trained under these settings, 
one can produce a sample from $q_{\sigma_L}(\Vec{x})$
via an annealed version of Langevin dynamics with 
a special step size schedule such that 
$\gamma_l = \varepsilon \cdot \sigma_l^2/\sigma_L^2$ 
for the $l$th noise level, where
$\varepsilon$ is a scaling factor (\refalgo{langevin}).

\begin{algorithm}[t!]
\caption{Annealed Langevin dynamics \cite{Song2019}}
\label{algo:langevin}
\begin{algorithmic}
\REQUIRE $\{\sigma_l\}_{l=1}^{L}$, $\varepsilon$
\STATE Initialize $\Vec{x} \sim \mathcal{N}(\Vec{0},\Vec{I})$
\FOR{$l=1$ to $L$}
\STATE $\gamma_l\leftarrow \varepsilon\cdot \sigma_l^2/\sigma_L^2$
\FOR{$t=1$ to $T$}
\STATE Draw $\Vec{z} \sim\mathcal{N}(\Vec{0},\Vec{I})$
\STATE Update $\Vec{x} \leftarrow \Vec{x} + \gamma_l 
\Vec{s}_{\theta}(\Vec{x},l)
+\sqrt{2\gamma_l} \Vec{z}$
\ENDFOR
\ENDFOR
\RETURN $\Vec{x}$
\end{algorithmic}
\end{algorithm}

\section{Formulation as Diffusion Models}

As revealed by Ho et al. \cite{Ho2020}, DSM is closely related to diffusion probabilistic models (DPMs), or diffusion models for short. 
Here, we review the principle of DPMs and show that their training objective and sample generation process have similarities to those of DSM described above.

Given a data sample $\Vec{x}_0$, normalized to have mean zero and unit variance, the diffusion process of DPM is defined as a Markov chain that gradually adds Gaussian noise to $\Vec{x}_0$. 
Based on this assumed process, the model, parametrized by $\theta$, is trained to find a reverse process that gradually reconstructs $\Vec{x}_0$ from its diffused versions.
Formally, DPMs are latent variable models of the form 
\begin{align}
p_{\theta}(\Vec{x}_0) = 
\int
p_{\theta}(\Vec{x}_{0:L})
\mbox{d}
\Vec{x}_{1:L},
\end{align}
where the notation $\Vec{x}_{i:j}$ is used to denote the set $\{\Vec{x}_{i},\ldots,\Vec{x}_{j}\}$ and 
$\Vec{x}_{1},\ldots,\Vec{x}_{L}$ are latent variables corresponding to the diffused versions of $\Vec{x}_0$, with $l=1,\ldots,L$ being the time step of the diffusion process.
The joint distribution $q(\Vec{x}_{1:L}|\Vec{x}_0)$ given $\Vec{x}_0$ is called the diffusion process, which is assumed to be a Markov chain that factors as
\begin{align}
q(\Vec{x}_{1:L}|\Vec{x}_0) = \prod_{l=1}^{L} q(\Vec{x}_{l}|\Vec{x}_{l-1}),
\end{align}
where 
$q(\Vec{x}_{l}|\Vec{x}_{l-1})$ is defined as
\begin{align}
q(\Vec{x}_{l}|\Vec{x}_{l-1})=
\mathcal{N}(\Vec{x}_{l}; \sqrt{1-\beta_l}\Vec{x}_{l-1}, \beta_l\Vec{I}).
\label{eq:DPM_1}
\end{align}
\refeq{DPM_1} can be viewed as a process of scaling $\Vec{x}_{l-1}$ by $\sqrt{1-\beta_l}$ and then adding Gaussian noise with variance $\beta_l$.
This scaling has the role of keeping the variance of $\Vec{x}_l$ at unity at each time step.
It is important to note that $q(\Vec{x}_{l} | \Vec{x}_0)$ at any time step $l$ can be obtained analytically as
\begin{align}
q(\Vec{x}_{l} | \Vec{x}_0)
=
\mathcal{N}(
\Vec{x}_{l};
\sqrt{\bar{\alpha}_l}
\Vec{x}_0,
(1-\bar{\alpha}_l)
\Vec{I}
),
\label{eq:DPM_2}
\end{align}
where $\alpha_l = 1-\beta_l$ and $\bar{\alpha}_l = \prod_{i=1}^{l} \alpha_{i}$,
from the fact that the sum of Gaussian random variables is also Gaussian.
If we use $\Vec{\epsilon} \sim \mathcal{N}(\Vec{0}, \Vec{I})$ to denote a standard Gaussian random variable,
\refeq{DPM_2} can be rewritten as 
\begin{align}
\Vec{x}_l = \sqrt{\bar{\alpha}_l} \Vec{x}_0 + \sqrt{1-\bar{\alpha}_l} \Vec{\epsilon}.
\label{eq:DPM_3}
\end{align}

The joint distribution $p_{\theta}(\Vec{x}_{0:L})$ is called the reverse process, which is also assumed to be a Markov chain that factors as
\begin{align}
p_{\theta}(\Vec{x}_{0:L}) = 
p_{\theta}(\Vec{x}_L) \prod_{l=1}^{L} p_{\theta}(\Vec{x}_{l-1}|\Vec{x}_{l}),
\label{eq:DPM_4}
\end{align}
starting from
$\Vec{x}_L\sim\mathcal{N}(\Vec{0},\Vec{I})$, where
$p_{\theta}(\Vec{x}_{l-1}|\Vec{x}_{l})$ is defined as
\begin{align}
p_{\theta}(\Vec{x}_{l-1}|\Vec{x}_{l})=
\mathcal{N}(
\Vec{x}_{l-1}; 
\Vec{\mu}_{\theta}(\Vec{x}_l, l),
\nu_l^2 \Vec{I}
).
\label{eq:DPM_5}
\end{align}
Here, 
$\nu_l$ is an untrained time-dependent constant and
$\Vec{\mu}_{\theta}(\Vec{x}_l, l)$ 
is assumed to be the ouput of a deep neural network that is parameterized by $\theta$ and conditioned on $l$.
According to reference \cite{Ho2020}, setting $\nu^2_l$ to $\beta_l$ has been experimentally found to work well.
Now, we can use the variational bound on the negative log-likelihood $\mathbb{E}[-\log p_{\theta}(\Vec{x}_0)]$
as the training objective for $\theta$ to be minimized:
\begin{align}
\mathcal{L}_{\rm DPM}(\theta) = 
\mathbb{E}_{\Vec{x}_{1:L}\sim q(\Vec{x}_{1:L}|\Vec{x}_0)}
\left[
\log \frac{
q(\Vec{x}_{1:L}|\Vec{x}_0)
}{
p_{\theta}(\Vec{x}_{0:L})
}
\right].
\label{eq:DPM_6}
\end{align}
It is important to note that since the exact bound of $\mathcal{L}_{\rm DPM}(\theta)$ is achieved when the Kullback-Leibler (KL) divergence between $p_{\theta}(\Vec{x}_{1:L}|\Vec{x}_0)$ and $q(\Vec{x}_{1:L}|\Vec{x}_0)$ is 0, minimizing $\mathcal{L}_{\rm DPM}(\theta)$ with respect to $\theta$ means not only fitting $p_{\theta}$ to the data distribution, but also making $p_{\theta}$ and $q$ as consistent as possible.
By using \refeq{DPM_3} and
a reparameterization
\begin{align}
\Vec{\mu}_{\theta}(\Vec{x}_l, l)
=
\frac{1}{\sqrt{\alpha_l}}
\left(
\Vec{x}_l - 
\frac{
1 - \alpha_l
}{
\sqrt{1-\bar{\alpha}_l}
}
\Vec{\epsilon}_{\theta}(\Vec{x}_l, l)
\right),
\label{eq:DPM_7}
\end{align}
it can be shown that 
\refeq{DPM_6} can be rewritten as
\begin{multline}
\mathcal{L}_{\rm DPM}(\theta) = 
\\
\frac{1}{L}
\sum_{l=1}^{L}
\mathbb{E}_{\Vec{x}_0, \Vec{\epsilon}}
[
c_l
\|
\Vec{\epsilon}_{\theta}(
\sqrt{\bar{\alpha}_l} \Vec{x}_0 + \sqrt{1-\bar{\alpha}_l} \Vec{\epsilon},
l
)
-
\Vec{\epsilon}
\|_2^2
]
,
\label{eq:DPM_8}
\end{multline}
where the expectation is taken over  the training examples of $\Vec{x}_0$ and the random samples of $\Vec{\epsilon}\sim \mathcal{N}(\Vec{0},\Vec{I})$, and $c_l$ is a constant related to $\alpha_l$, $\bar{\alpha}_l$, and $\nu_l$.
See reference \cite{Ho2020} for the detailed derivation of \refeq{DPM_8}.
As it has been reported by Ho et al. that better model training can be achieved by setting $c_l$ to 1 \cite{Ho2020}, $c_l$ is henceforth set to 1.
The reparameterization of \refeq{DPM_7} implies representing $\Vec{\epsilon}_{\theta}(\Vec{x}_l, l)$ instead of $\Vec{\mu}_{\theta}(\Vec{x}_l, l)$ as a neural network.

Let us now compare \refeq{DPM_8} with \refeq{weighted_denoising_score_matching}.
By using the symbol $\Vec{x}_0$ instead of $\Vec{x}$ and
using a reparameterization $\tilde{\Vec{x}} = \Vec{x}_0 + \sigma_l \Vec{\epsilon}$, 
\refeq{weighted_denoising_score_matching} can be rewritten as 
\begin{align}
\mathcal{L}_{\rm DSM}(\theta) =
\frac{1}{L}
\sum_{l=1}^{L}
\mathbb{E}_{\Vec{x}_0,\Vec{\epsilon}}
[
\|
\sigma_l \Vec{s}_{\theta}(\Vec{x}_0 + \sigma_l \Vec{\epsilon}, l)
+
\Vec{\epsilon}
\|_2^2
],
\label{eq:DPM_9}
\end{align}
where the expectation is taken over  the training examples of $\Vec{x}_0$ and the random samples of $\Vec{\epsilon}\sim \mathcal{N}(\Vec{0},\Vec{I})$.
A comparison of \refeq{DPM_9} and \refeq{DPM_8} reveals that 
$-\sigma_l \Vec{s}_{\theta}(\Vec{x}_0 + \sigma_l \Vec{\epsilon}, l)$
and 
$\Vec{\epsilon}_{\theta}(\sqrt{\bar{\alpha}_l} \Vec{x}_0 + \sqrt{1-\bar{\alpha}_l} \Vec{\epsilon},l)$
are in correspondence with each other.
Hence, we will henceforth refer to $\Vec{\epsilon}_{\theta}$ as the score approximator as well.
The main difference is that the input to the network is
$\Vec{x}_0 + \sigma_l \Vec{\epsilon}$ in DSM, 
while it is
$\sqrt{\bar{\alpha}_l} \Vec{x}_0 + \sqrt{1-\bar{\alpha}_l} \Vec{\epsilon}$ in DPM.
This means that in DSM the variance of the network input $\Vec{x}_l$ is assumed to increase after the addition of Gaussian noise, whereas in DPM the variance is assumed to remain 1 owing to the scaling of $\Vec{x}_0$.

\begin{algorithm}[t!]
\caption{Reverse diffusion process}
\label{algo:DPM}
\begin{algorithmic}
\REQUIRE $\{\alpha_l\}_{l=1}^{L}$
\STATE Initialize $\Vec{x} \sim \mathcal{N}(\Vec{0},\Vec{I})$
\FOR{$l=L$ to $1$}
\STATE Draw $\Vec{z} \sim\mathcal{N}(\Vec{0},\Vec{I})$
\STATE Update 
$\Vec{x}
\leftarrow
\frac{1}{\sqrt{\alpha_l}}
\left(
\Vec{x} - 
\frac{
1 - \alpha_l
}{
\sqrt{1-\bar{\alpha}_l}
}
\Vec{\epsilon}_{\theta}(\Vec{x}, l)
\right) + \nu_l \Vec{z}$
\ENDFOR
\RETURN $\Vec{x}$
\end{algorithmic}
\end{algorithm}

Once $\theta$ is trained, one can produce a sample from $p_{\theta}(\Vec{x})$ in accordance with the Markov chain of the reverse process (\refalgo{DPM}).
A comparison of \refalgos{langevin}{DPM} reveals the connection between the sample generation algorithms 
based on the Langevin dynamics and the reverse diffusion process.
Notice that in \refalgo{langevin}, the loop counter $l$ increments from $1$ to $L$ in the for-loop, whereas in \refalgo{DPM} it decrements from $L$ to $1$. This is just a matter of how the diffusion time step or noise level is indexed, not an essential difference. 
In fact, if we define $l' = L - l + 1$ as the new loop counter in \refalgo{langevin}, we can rewrite  \refalgo{langevin} to decrement the loop counter $l'$ from $L$ to $1$, as in \refalgo{DPM}.
Let us now focus on the update equations in \refalgos{langevin}{DPM}.
Since $\nabla_{\Vec{x}_l}\log q(\Vec{x}_l) = \mathbb{E}_{\Vec{x}_0}[\nabla_{\Vec{x}_l}\log q(\Vec{x}_l|\Vec{x}_0)]$ (see Appendix for the proof) and $\nabla_{\Vec{x}_l}\log q(\Vec{x}_l|\Vec{x}_0) = - \frac{\Vec{\epsilon}}{\sqrt{1-\bar{\alpha}_l}}$, we have
\begin{align}
\nabla_{\Vec{x}_l}\log q(\Vec{x}_l) =
- \frac{\Vec{\epsilon}}{\sqrt{1-\bar{\alpha}_l}},
\label{eq:DPM_10}
\end{align}
where $\Vec{\epsilon} \sim \mathcal{N}(\Vec{0},\Vec{I})$.
Since $\Vec{\epsilon}_{\theta}(\Vec{x}_l, l)$ is trained to predict $\Vec{\epsilon}$ in \refeq{DPM_10}, if $\theta$ is successfully trained, $- \frac{\Vec{\epsilon}_{\theta}(\Vec{x}_l, l)}{\sqrt{1-\bar{\alpha}_l}}$ should be a good approximation of $\nabla_{\Vec{x}_l}\log q(\Vec{x}_l)$.
Therefore, the update equation in \refalgo{DPM} can be interpreted as moving $\Vec{x}_l$ in the direction of  $\nabla_{\Vec{x}_l}\log q(\Vec{x}_l)$
with step size $1-\alpha_l$, then scaling by $\frac{1}{\sqrt{\alpha}_l}$, and finally adding $\nu_l\Vec{z}$. 
Thus, this can be seen as the same process as the update equation in \refalgo{langevin}, except for the scaling.
This scaling corresponds to the inverse of the scaling for $\Vec{x}_{l-1}$ assumed in \refeq{DPM_1}. 
Recall that 
$-\sigma_l \Vec{s}_{\theta}(\Vec{x}_l,l)$ in DSM corresponds to 
$\Vec{\epsilon}_{\theta}(\Vec{x}_l,l)$ in DPM. 
Since $\Vec{s}_{\theta}(\Vec{x}_l,l)$ is trained to approximate $\nabla_{\Vec{x}_l}\log q(\Vec{x}_l)$, 
from \refeq{DPM_10}, this implies that $\sigma_l$ corresponds to $\sqrt{1-\bar{\alpha}_l}$. 

\section{Related Work}
\label{sec:relatedwork}

As the name implies, 
VoiceGrad is inspired by WaveGrad \cite{Chen2020}, 
a recently introduced novel neural waveform generation method
based on the concept of 
DPMs.
The idea 
is to 
model the process of
gradually converting a Gaussian white noise signal
into a speech waveform that is best associated with a conditioning mel-spectrogram as the reverse diffusion process. 
After training,
one can generate 
a waveform (in a non-autoregressive manner, unlike WaveNet)
given a mel-spectrogram 
via 
the trained reverse diffusion process, 
starting from a randomly drawn noise signal.
One straightforward way of adapting this idea to VC tasks would be to
use the model to generate
the acoustic feature sequence of target speech
and that of source speech as the conditioning input. 
This idea may work if time-aligned parallel utterances of 
a speaker pair are available,
but here we are concerned with 
achieving non-parallel any-to-many VC, as described below.

Note that several DPM-based VC methods, such as Diff-SVC \cite{Liu2021b} and Diff-VC \cite{Popov2022}, have been proposed after the publication of the first version of the preprint paper on this work \cite{Kameoka2020arXiv_VoiceGrad}.
The idea of Diff-VC is to first convert an input mel-spectrogram to an ``average voice'' mel-spectrogram using an encoder trained by phoneme supervision on speech samples from many speakers, and then to a target speaker mel-spectrogram using the reverse diffusion process of a trained DPM. 
As described below, our VoiceGrad differs in that it performs the Langevin dynamics or reverse diffusion process starting directly from the mel-spectrogram of source speech. 

\section{VoiceGrad}

\subsection{Key Idea}
\label{subsec:key_idea}

\begin{figure*}[t!]
\centering
\begin{minipage}[t!]{.65\linewidth}
  \centerline{\includegraphics[width=.98\linewidth]{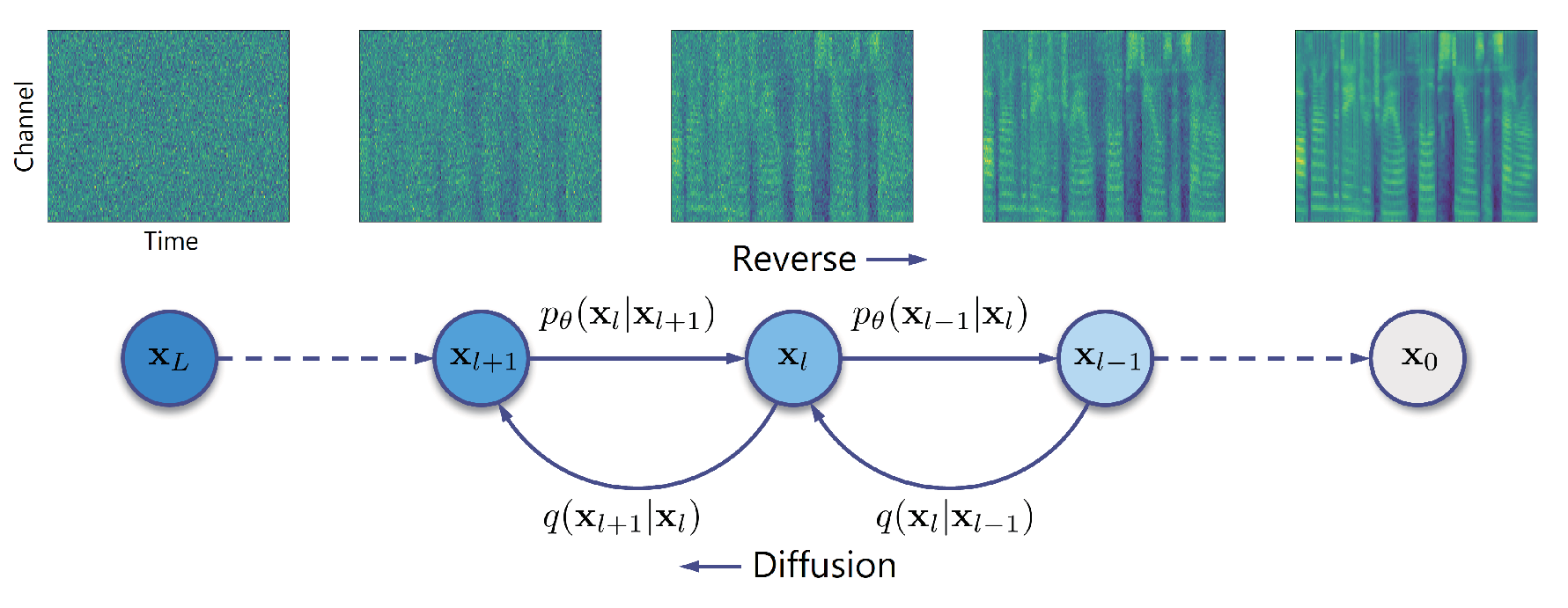}}
  \vspace{-2ex}
  \caption{Mel-spectrogram generation by reverse diffusion process}
\label{fig:dpm}
\end{minipage}
\end{figure*}

We will now describe how
VoiceGrad uses the concepts of  
DSM, 
Langevin dynamics, 
and DPMs 
to achieve non-parallel any-to-many VC. 
Given 
a source speech feature sequence (e.g., mel-spectrogram),
our key idea is to 
formulate the VC problem as 
finding the stationary point of the
log density of target speech feature sequences 
nearest to the source sequence.
Thus, we can naturally think of employing the Langevin dynamics or reverse diffusion process starting from a certain noise level 
to perform VC by using the source speech feature sequence as an initial point
and moving it along the gradient direction of the log density of 
target speech feature sequences. 
From the DPM perspective, this corresponds to assuming that the source speech feature sequence is a diffused version of a target speech feature sequence.
Although this process does not 
necessarily ensure the preservation of the linguistic content in source speech,
it was experimentally found to work under some settings, as detailed later.
To enable a single score approximator to predict the score functions of 
the feature sequences of all the $K$ target speakers included in the training set, 
we also condition the score approximator network on the target speaker index $k\in\{1,\ldots,K\}$.
Namely, VoiceGrad
uses a network $\Vec{\epsilon}_{\theta}(\Vec{x},l,k)$
to describe the score approximator, where $\Vec{x}$ denotes the input speech feature.

Owing to the idea described above, 
VoiceGrad does not require the training set to consist of parallel utterances, 
allows the speaker of input speech at test time to be arbitrary, 
and
can convert input speech to the voices of multiple known speakers using a single trained network.

\subsection{Acoustic Feature and Waveform Generation}

We use the 80-dimensional log mel-spectrogram extracted from input speech as the acoustic feature sequence to be converted, and choose to use HiFi-GAN \cite{Kong2020} for waveform generation from the converted mel-spectrogram.
At training time, each element $x_{d,m}$ of the log mel-spectrogram $\Vec{x}$ of each training example is normalized to $x_{d,m} \leftarrow (x_{d,m} - \psi_d)/\zeta_d$, 
where $d$ denotes the mel-filterbank channel of the mel-spectrogram,
$m$ denotes the frame index, 
and $\psi_d$ and $\zeta_d$ denote 
the mean and standard deviation of the $d$-th channel elements of all the training samples.
At test time, the mel-spectrogram of input speech is normalized in the same way.

\subsection{Training and Conversion Processes}

As noted above, in VoiceGrad, the generation process of a target speaker's mel-spectrogram is treated as a gradient descent search for a probabilistically likely mel-spectrogram in the DSM version, whereas in the DPM version, it is regarded as the reverse diffusion process (\reffig{dpm}).

As in the previous study \cite{Chen2020}, using the $L_1$ measure instead of the $L_2$ measure in \refeq{weighted_denoising_score_matching} and \refeq{DPM_8} was found to be effective in terms of both training stability and audio quality at test time. 
Hence, the training objectives for VoiceGrad
to be minimized with respect to $\theta$
under the DSM and DPM formulations 
become
\begin{align}
\mathcal{L}_{\rm DSM}(\theta) &=
\mathbb{E}
[
\|
\Vec{\epsilon}_{\theta}(\Vec{x}_0 + \sigma_l \Vec{\epsilon}, l, k)
-
\Vec{\epsilon}
\|_1
],
\label{eq:VoiceGrad_Training_Objective_DSM}
\\
\mathcal{L}_{\rm DPM}(\theta) &= 
\mathbb{E}
[
\|
\Vec{\epsilon}_{\theta}(
\sqrt{\bar{\alpha}_l} \Vec{x}_0 + \sqrt{1-\bar{\alpha}_l} \Vec{\epsilon},
l, k
)
-
\Vec{\epsilon}
\|_1
]
,
\label{eq:VoiceGrad_Training_Objective_DPM}
\end{align}
respectively, 
where 
the expectations in both equations are taken over 
the random samples of 
$l\sim \mathbb{U}(1,\ldots,L)$,
$k\sim \mathbb{U}(1,\ldots,K)$, 
and
$\Vec{\epsilon} \sim \mathcal{N}(\Vec{0},\Vec{I})$, and 
the training examples of 
$\Vec{x}_0 \sim p(\Vec{x}_0|k)$.
Here, $\mathbb{U}(\cdot)$ is used to denote a discrete uniform distribution over the integers in its argument.
Note that $-\sigma_l \Vec{s}_{\theta}(\Vec{x}_l, l, k)$ is expressed here as $\Vec{\epsilon}_{\theta}(\Vec{x}_l, l, k)$ in \refeq{VoiceGrad_Training_Objective_DSM} to unify the symbols for the score approximator.
Given a set of training examples 
$\mathcal{X}=\{\Vec{x}_0^{(k,n)}\}_{1\le k\le K, 1\le n\le N}$, 
where 
$N$ is the number of training utterances of each speaker,
$\Vec{x}_0^{(k,n)}\in \mathbb{R}^{80\times M_{k,n}}$ 
denote 
the mel-spectrogram of the $n$th training utterance of the $k$th speaker, respectively, 
$M_{k,n}$ denotes the length of $\Vec{x}_0^{(k,n)}$,
the expectation $\mathbb{E}_{k,\Vec{x}_0}[\cdot]$
can be approximated as 
the sample mean over $\mathcal{X}$, 
and 
$\mathbb{E}_{\Vec{\epsilon}}[\cdot]$ can be 
evaluated using the Monte Carlo approximation. 

\begin{figure*}[t!]
\centering
\begin{minipage}[t]{.6\linewidth}
  \centerline{\includegraphics[width=.98\linewidth]{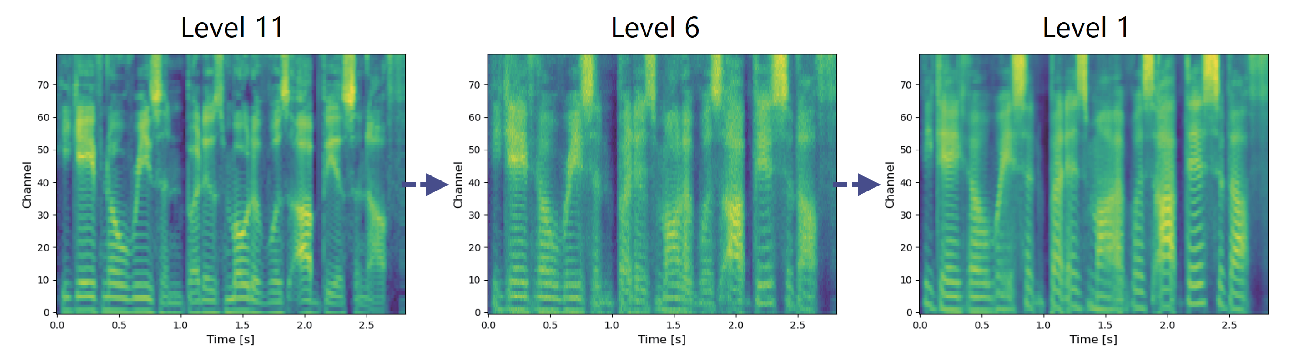}}
  \vspace{-2ex}
  \caption{Process of converting a male speaker's mel-spectrogram to a female voice by a reverse diffusion process starting from level 11 at test time.}
  \label{fig:conversion_process}
  \end{minipage}
\end{figure*}

Once the score approximator $\Vec{\epsilon}_{\theta}$ is trained using \refeq{VoiceGrad_Training_Objective_DSM} or \refeq{VoiceGrad_Training_Objective_DPM} as a criterion, we
can use \refalgo{voicegrad_dsm} 
or 
\refalgo{voicegrad_dpm}
to convert an input mel-spectrogram $\Vec{x}$ to the voice of speaker $k$.
For both algorithms, we found experimentally that starting the iteration from a certain point in the middle rather than from the beginning is effective in terms of the audio quality of the converted speech.
Henceforth, we use $L'$ to denote the starting noise level of the iteration.
\reffig{conversion_process} shows an example of the actual process of converting a male speaker's mel-spectrogram to a female voice by reverse diffusion process (\refalgo{voicegrad_dpm}) starting from $L'=11$.


\begin{algorithm}[t]
\caption{DSM-based VoiceGrad}
\label{algo:voicegrad_dsm}
\begin{algorithmic}
\REQUIRE $\{\sigma_l\}_{l=L'}^{L}$, $\varepsilon$, $T$, $\Vec{x}$, $k$
\FOR{$l=L'$ to $L$}
\STATE $\gamma_l\leftarrow \varepsilon\cdot \sigma_l^2/\sigma_L^2$
\FOR{$t=1$ to $T$}
\STATE Draw $\Vec{z} \sim\mathcal{N}(\Vec{0},\Vec{I})$
\STATE Update $\Vec{x} \leftarrow \Vec{x} 
- \frac{\gamma_l}{\sigma_l} \Vec{\epsilon}_{\theta}(\Vec{x}, l, k)
+\sqrt{2\gamma_l} \Vec{z}$
\ENDFOR
\ENDFOR
\RETURN $\Vec{x}$
\end{algorithmic}
\end{algorithm}

\begin{algorithm}[t]
\caption{DPM-based VoiceGrad}
\label{algo:voicegrad_dpm}
\begin{algorithmic}
\REQUIRE $\{\alpha_l\}_{l=1}^{L'}$, $\{\bar{\alpha}_l\}_{l=1}^{L'}$, $\Vec{x}$, $k$
\FOR{$l=L'$ to $1$}
\STATE Draw $\Vec{z} \sim\mathcal{N}(\Vec{0},\Vec{I})$
\STATE Update 
$\Vec{x}
\leftarrow
\frac{1}{\sqrt{\alpha_l}}
\left(
\Vec{x} - 
\frac{
1 - \alpha_l
}{
\sqrt{1-\bar{\alpha}_l}
}
\Vec{\epsilon}_{\theta}(\Vec{x}, l, k)
\right) + \nu_l \Vec{z}$
\ENDFOR
\RETURN $\Vec{x}$
\end{algorithmic}
\end{algorithm}

\subsection{BNF Conditioning}
\label{subsec:bnf_conditioning}

In \refsubsec{key_idea}, we noted  that the Langevin dynamics or reverse diffusion process, starting from the mel-spectrogram of source speech, has been experimentally found to preserve the linguistic content in the source speech to a certain extent.
However, 
we have also found that by designing the score approximator to incorporate 
the 
BNF
sequence obtained by an ASR model from source speech, the conversion process can be guided to preserve the linguistic content and improve the intelligibility of the generated speech. 

To extract the  BNF sequence from a mel-spectrogram, we use the bottleneck feature extractor (BNE) proposed by Liu et al. \cite{Liu2021}.
This BNE is obtained by inserting an additional bottleneck layer between the encoder and decoder in an end-to-end phoneme recognizer \cite{Kim2017}, training the entire network on a large speech recognition data corpus, and dropping all the layers subsequent to the bottleneck layer from the network after training.
For more details on the architecture and training scheme of the BNE, see reference \cite{Liu2021}.

The BNF sequence is expected to contain little information other than the linguistic content of the input speech.
Namely, the BNF sequence should remain virtually unchanged before and after VC.
Therefore, it is expected to work even if the BNF sequence of target speech is used as an additional input to the network during the score approximator training whereas the BNF sequence of the source speech is used instead at test time.
By using the BNF sequence in this way, 
we expect that the BNF sequence will encourage the score approximator to learn to predict the target score function using the linguistic information of the input speech as a guide.
With the above expectation, the BNF sequence is obtained from the mel-spectrogram of each training example at training time and from the mel-spectrogram of source speech at test time.

The current model setup is such that the output of the trained BNE becomes a sequence of 144-dimensional BNF vectors of the same length as the mel-spectrogram input to the BNE. 


\subsection{Noise Variance Scheduling}
\label{subsec:scheduling}

Although the noise variances, i.e., $\{\sigma_l\}_l$ in the DSM formulation and $\{\beta_l\}_l$ in the DPM formulation, can be set arbitrarily, it has been experimentally reported that the choice of these variances can affect sample quality in image generation applications. 
For the DSM formulation, we set
$\{\sigma_l\}_{1\le l\le L}$ 
at a geometric sequence, as in \cite{Song2019}, with common ratio
$\frac{\sigma_2}{\sigma_1}=\cdots=\frac{\sigma_{L}}{\sigma_{L-1}}\approx 0.787$,
where $L=21$, $\sigma_1=1.2$, and $\sigma_L = 0.01$.
For the DPM formulation, we use a cosine-based schedule \cite{Nichol2021} for the noise variance setting.
Specifically, we construct a schedule in terms of $\bar{\alpha}_l$ (instead of $\beta_l$) as
\begin{align}
\bar{\alpha}_l = \frac{f(l)}{f(0)},~~
f(l) = \cos
\left(
\frac{l/L + \eta}{1 + \eta}
\cdot
\frac{\pi}{2}
\right)^2,
\end{align}
where $L=20$.
From the relation between $\bar{\alpha}_l$ and $\beta_l$, we get $\beta_l=1-\frac{\bar{\alpha}_l}{\bar{\alpha}_{l-1}}$.
To prevent $\beta_l$ from being too close to 1, 
$\beta_l$ is further clipped to be no larger than 0.999.
$\eta$ is a small offset to prevent $\beta_l$ from being too small when close to $l=0$, which we set at $\eta=0.008$ in the following experiment. 

\subsection{Network Architecture}
\label{subsec:netarch}

\subsubsection{U-Net-like Structure}

\begin{figure}[t!]
\centering
  \centerline{\includegraphics[width=.98\linewidth]{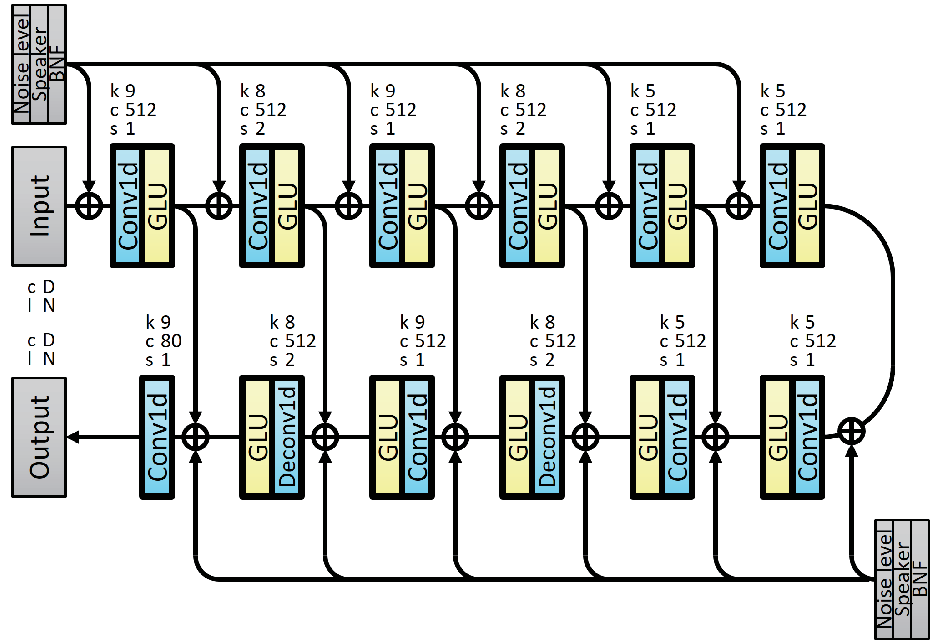}}
  \vspace{-0ex}
  \caption{Network architecture of the score approximator with a U-Net-like fully convolutional structure. Here, $\oplus$ represents array concatenation along the channel direction. See \refsubsubsec{conditioning} for details on how the noise-level and speaker indices and the BNF sequence are incorporated into the network.}
\label{fig:netarch}
\end{figure}

The architecture of the score approximator 
is detailed in \reffig{netarch}.
As \reffig{netarch} shows, it is designed to 
have a fully convolutional
structure similar to U-Net \cite{Ronneberger2015} that
takes the mel-spectrogram of input speech as 
an input array 
and outputs an equally sized array.
Note that we have also tried other types of architectures, such as AE-like bottleneck architectures without skip connections, but so far we have experimentally confirmed that the current U-Net-like architecture works best.
Here, the input and output of each layer are vector sequences, where ``c'' and ``l'' denote the channel number and the length of a vector sequence, respectively. ``Conv1d'', ``GLU'', and ``Deconv1d'' denote 1D convolution, gated linear unit (GLU) \cite{Dauphin2017short}, and 1D transposed convolution layers, respectively.
See below for the details of GLUs. 
``k'', ``c'' and ``s'' denote the kernel size, output channel number, and stride size of a convolution layer, respectively.
All the convolution weights are initilized using the Glorot normal initializer \cite{Glorot2010} with gain 0.5 and reparameterized using weight normalization \cite{Salimans2016}. 


\subsubsection{Gated Linear Unit}

For non-linear activation functions, we use GLUs \cite{Dauphin2017short}. 
The output of a GLU is defined as 
$\mathsf{GLU}(\Vec{y}) = \Vec{y}_1 \odot \mathsf{sigmoid}(\Vec{y}_2)$ 
where $\Vec{y}$ is the input, 
$\Vec{y}_1$ and $\Vec{y}_2$ are equally sized arrays 
obtained by splitting $\Vec{y}$ along the channel dimension,
and $\mathsf{sigmoid}$ is a sigmoid gate function.
Like long short-term memory units, 
GLUs provide a linear path for
the gradients while retaining non-linear capabilities,
thus reducing the vanishing gradient
problem for deep architectures.

\subsubsection{Noise-level, Speaker, and BNF Conditioning}
\label{subsubsec:conditioning}

The noise-level and speaker indices are incorporated into each convolution layer in 
the score approximator
by first retrieving embedding vectors from two learnable lookup tables according to the specified indices, 
then repeating those vectors in the time direction to the length compatible with the input of the convolution layer, 
and finally concatenating the two repeated vector sequences to the input along the channel direction. 

In the version that incorporates a BNF sequence $\Vec{p}$ into the network, 
$\Vec{p}$ is 
first fed into a strided convolution layer $h$ with 32 output channels with a stride size $r$,
and then the output is
appended along the channel direction to the input of each convolution layer with GLU.
The stride size $r$ is appropriately chosen so that the length of the output of $h$ is compatible with the input of the convolution layer with GLU.


\section{Experiments}
\label{sec:experiments}

\subsection{Dataset}
\label{subsec:dataset}

To evaluate the performance of VoiceGrad, 
we conducted speaker conversion experiments.
For the experiments, we used the CMU ARCTIC database \cite{Kominek2004short},
which consists of recordings of 
18 speakers
each reading the same 1,132 phonetically balanced English sentences.
For the training set and the test set for a closed-set scenario, 
we used the utterances of
two female speakers, `clb' and `slt', 
and two male speakers, `bdl' and `rms'. 
Thus, $K=4$.
We also used the utterances of 
two male speakers, `jmk' and `ksp', and
a female speaker, `lnh', as the test set for an open-set scenario.
All the speech signals were sampled at 16 kHz. 

For each speaker,
we first split the 1,132 sentences into
1,000, 100, and 32 sentences and
used the 32 sentences for evaluation.
To simulate a non-parallel training scenario,
we further
divided the 1,000 sentences equally
into four groups and used 
the first, second, third, and fourth groups for training
for speakers clb, bdl, slt, and rms, respectively,
so as not to use 
the same sentences between different speakers.
The training utterances of speakers clb, bdl, slt, and rms
were about 12, 11, 11, and 14 minutes long in total, respectively.
For the test set, we used the test utterances of 
speakers clb, bdl, slt, and rms for the closed-set scenario,
and those of speakers jmk, ksp and lnh for the open-set scenario.

\subsection{Baseline Methods}

We chose 
AE-based zero-shot VC method \cite{Qian2019} (AutoVC),
PPG-based one-shot VC method \cite{Liu2021} (PPG-VC),
and 
our previously proposed StarGAN-VC \cite{Kameoka2018SLTshort_StarGAN-VC, Kameoka2020IEEE-ASLP_StarGAN} 
for comparison,
as 
these methods, in principle, are capable of addressing many-to-many scenarios but also 
any-to-many situations by leveraging non-parallel corpora.
It should also be noted that these methods are employed in the systems submitted for VCC2020.
To run these methods, we used the source codes provided by the respective authors \cite{autovc_github, ppgvc_github, starganvc_github}. 
For StarGAN-VC, we used the Wasserstein distance as the training objective \cite{Arjovsky2017, Gulrajani2017, Choi2018, Kameoka2020IEEE-ASLP_StarGAN}. 
Although in the original paper \cite{Kameoka2020IEEE-ASLP_StarGAN}, StarGAN-VC used the mel-cepstral coefficient (MCC) vector sequence of source speech as the feature to be converted, in this experiment it was modified to use the mel-spectrogram instead as the feature to be converted and use HiFi-GAN \cite{Kong2020} to generate waveforms from the converted spectrograms \cite{starganvc_github,starganvc_url}.

\subsection{Model Setup}

The
score approximator
network 
was trained
using 
the Adam optimizer \cite{Kingma2015short} 
with random initialization,
the learning rate of 0.001,
and 
the mini-batch size of 16.
The starting noise level index $L'$ was set to 
4 for \refalgo{voicegrad_dsm} and 
11 for \refalgo{voicegrad_dpm}.
The noise variances 
$\{\sigma_l\}_{1\le l\le L}$ and $\{\beta_l\}_{1\le l \le L}$ were set as described in \refsubsec{scheduling}.
The step size parameter $\varepsilon$ 
and the iteration number $T$
in \refalgo{voicegrad_dsm} were set at $10^{-5}$ and 
32,
respectively.
Thus, the number of iterations required to complete the conversion was 
$(21-4+1)\times 32=576$ for \refalgo{voicegrad_dsm} and 
$11$ for \refalgo{voicegrad_dpm}.
In \refalgo{voicegrad_dpm}, $\nu_l$ is set to $\sqrt{\beta_l}$, following the guidance in reference \cite{Ho2020}.



\begin{table*}[t!]
\caption{Comparisons of the DSM and DPM Versions}
\label{tab:dsm_dpm_comp}
\vspace{-2ex}
\centering
\begin{scriptsize}
\begin{tabular}{l|l V{3} c|c|c|c|c|c|c|c}
\thline
\multicolumn{2}{c V{3}}{Speakers}&
\multicolumn{2}{c|}{$\downarrow$MCD [dB]}&
\multicolumn{2}{c|}{$\uparrow$LFC}&
\multicolumn{2}{c|}{$\downarrow$CER [\%]}&
\multicolumn{2}{c}{$\uparrow$pMOS}
\\\hline
\multicolumn{1}{c|}{s}&\multicolumn{1}{c V{3}}{t}&DSM&DPM&DSM&DPM&DSM&DPM&DSM&DPM\\\thline
      &   bdl&$9.06\pm .12$&$\bm{8.25\pm .12}$&$0.13\pm .05$&$\bm{0.38\pm .05}$&$14.6$&$\bm{13.7}$&$\bm{3.21\pm .05}$&$3.16\pm .05$\\
   clb&   slt&$7.24\pm .07$&$\bm{6.34\pm .07}$&$0.49\pm .04$&$\bm{0.71\pm .03}$&$\bm{7.1}$&$7.2$&$3.56\pm .05$&$\bm{3.78\pm .06}$\\
      &   rms&$8.45\pm .08$&$\bm{7.49\pm .08}$&$\bm{0.33\pm .05}$&$0.10\pm .05$&$15.8$&$\bm{12.8}$&$\bm{3.50\pm .05}$&$3.00\pm .05$\\\hline
      &   clb&$8.15\pm .11$&$\bm{7.08\pm .11}$&$0.14\pm .05$&$\bm{0.17\pm .04}$&$\bm{20.7}$&$22.8$&$2.76\pm .05$&$\bm{3.04\pm .06}$\\
   bdl&   slt&$7.75\pm .10$&$\bm{7.18\pm .09}$&$0.28\pm .04$&$\bm{0.41\pm .04}$&$\bm{16.3}$&$16.9$&$3.16\pm .06$&$\bm{3.43\pm .05}$\\
      &   rms&$8.36\pm .12$&$\bm{7.33\pm .14}$&$0.29\pm .05$&$\bm{0.44\pm .04}$&$\bm{10.8}$&$15.6$&$3.69\pm .04$&$\bm{3.77\pm .05}$\\\hline
      &   clb&$7.07\pm .08$&$\bm{6.24\pm .08}$&$0.46\pm .04$&$\bm{0.69\pm .03}$&$\bm{10.7}$&$11.6$&$3.23\pm .05$&$\bm{3.70\pm .05}$\\
   slt&   bdl&$9.00\pm .11$&$\bm{7.93\pm .12}$&$0.22\pm .05$&$\bm{0.55\pm .04}$&$15.0$&$\bm{10.1}$&$3.09\pm .06$&$\bm{3.23\pm .05}$\\
      &   rms&$8.56\pm .10$&$\bm{7.80\pm .11}$&$\bm{0.28\pm .05}$&$0.20\pm .05$&$17.2$&$\bm{12.1}$&$\bm{3.49\pm .05}$&$2.92\pm .05$\\\hline
      &   clb&$7.85\pm .08$&$\bm{7.05\pm .08}$&$\bm{0.27\pm .05}$&$0.07\pm .05$&$19.1$&$\bm{14.8}$&$2.90\pm .06$&$\bm{3.10\pm .07}$\\
   rms&   bdl&$9.00\pm .13$&$\bm{7.95\pm .17}$&$0.27\pm .06$&$\bm{0.29\pm .05}$&$\bm{5.9}$&$9.8$&$3.56\pm .04$&$\bm{3.72\pm .05}$\\
      &   slt&$7.84\pm .12$&$\bm{7.51\pm .11}$&$\bm{0.31\pm .05}$&$0.08\pm .05$&$17.0$&$\bm{13.7}$&$\bm{3.29\pm .05}$&$3.27\pm .05$\\\hline
\multicolumn{2}{c V{3}}{All pairs}
             &$8.19\pm .04$&$\bm{7.35\pm .04}$&$0.29\pm .01$&$\bm{0.34\pm .02}$&$14.2$&$\bm{13.4}$&$3.29\pm .02$&$\bm{3.34\pm .02}$\\\thline   
\end{tabular}
\end{scriptsize}
\end{table*}

\begin{table*}[t!]
\caption{Effect of BNF Conditioning in DSM Version}
\label{tab:bnf_conditioning_comp_dsm}
\vspace{-2ex}
\centering
\begin{scriptsize}
\begin{tabular}{l|l V{3} c|c|c|c|c|c|c|c}
\thline
\multicolumn{2}{c V{3}}{Speakers}&
\multicolumn{2}{c|}{$\downarrow$MCD [dB]}&
\multicolumn{2}{c|}{$\uparrow$LFC}&
\multicolumn{2}{c|}{$\downarrow$CER [\%]}&
\multicolumn{2}{c}{$\uparrow$pMOS}
\\\hline
\multicolumn{1}{c|}{s}&\multicolumn{1}{c V{3}}{t}&DSM&DSM+BNF&DSM&DSM+BNF&DSM&\!\!\!DSM+BNF\!\!\!&DSM&DSM+BNF\\\thline
      &   bdl&$9.06\pm .12$&$\bm{7.44\pm .09}$&$0.13\pm .05$&$\bm{0.53\pm .03}$&$14.6$&$\bm{2.5}$&$3.21\pm .05$&$\bm{3.61\pm .05}$\\
   clb&   slt&$7.24\pm .07$&$\bm{6.24\pm .06}$&$0.49\pm .04$&$\bm{0.71\pm .02}$&$7.1$&$\bm{2.3}$&$3.56\pm .06$&$\bm{3.69\pm .05}$\\
      &   rms&$8.45\pm .08$&$\bm{6.64\pm .07}$&$0.33\pm .05$&$\bm{0.57\pm .03}$&$15.8$&$\bm{2.7}$&$3.50\pm .05$&$\bm{3.71\pm .05}$\\\hline
      &   clb&$8.15\pm .11$&$\bm{6.72\pm .09}$&$0.14\pm .05$&$\bm{0.44\pm .04}$&$20.7$&$\bm{3.3}$&$2.76\pm .05$&$\bm{3.31\pm .06}$\\
   bdl&   slt&$7.75\pm .10$&$\bm{6.88\pm .07}$&$0.28\pm .04$&$\bm{0.61\pm .03}$&$16.3$&$\bm{3.1}$&$3.16\pm .06$&$\bm{3.62\pm .05}$\\
      &   rms&$8.36\pm .12$&$\bm{7.33\pm .14}$&$0.29\pm .05$&$\bm{0.43\pm .04}$&$10.8$&$\bm{2.9}$&$3.69\pm .04$&$\bm{3.81\pm .05}$\\\hline
      &   clb&$7.07\pm .08$&$\bm{6.14\pm .07}$&$0.46\pm .04$&$\bm{0.68\pm .03}$&$10.7$&$\bm{1.8}$&$3.23\pm .05$&$\bm{3.62\pm .05}$\\
   slt&   bdl&$9.00\pm .11$&$\bm{7.38\pm .08}$&$0.22\pm .05$&$\bm{0.57\pm .03}$&$15.0$&$\bm{2.2}$&$3.09\pm .06$&$\bm{3.68\pm .05}$\\
      &   rms&$8.56\pm .10$&$\bm{7.02\pm .11}$&$0.28\pm .05$&$\bm{0.50\pm .04}$&$17.2$&$\bm{1.8}$&$3.49\pm .05$&$\bm{3.65\pm .04}$\\\hline
      &   clb&$7.85\pm .08$&$\bm{6.45\pm .06}$&$0.27\pm .05$&$\bm{0.54\pm .03}$&$19.1$&$\bm{1.1}$&$2.90\pm .06$&$\bm{3.62\pm .06}$\\
   rms&   bdl&$9.00\pm .13$&$\bm{7.89\pm .16}$&$0.27\pm .06$&$\bm{0.42\pm .05}$&$5.9$&$\bm{1.2}$&$3.56\pm .04$&$\bm{3.80\pm .05}$\\
      &   slt&$7.84\pm .12$&$\bm{6.74\pm .09}$&$0.31\pm .05$&$\bm{0.55\pm .04}$&$17.0$&$\bm{1.4}$&$3.29\pm .05$&$\bm{3.79\pm .05}$\\\hline
\multicolumn{2}{c V{3}}{All pairs}
             &$8.19\pm .04$&$\bm{6.91\pm .04}$&$0.29\pm .01$&$\bm{0.55\pm .01}$&$14.2$&$\bm{2.2}$&$3.29\pm .02$&$\bm{3.66\pm .02}$\\\thline   
\end{tabular}
\end{scriptsize}
\end{table*}

\begin{table*}[t!]
\caption{Effect of BNF Conditioning in DPM Version}
\label{tab:bnf_conditioning_comp_dpm}
\vspace{-2ex}
\centering
\begin{scriptsize}
\begin{tabular}{l|l V{3} c|c|c|c|c|c|c|c}
\thline
\multicolumn{2}{c V{3}}{Speakers}&
\multicolumn{2}{c|}{$\downarrow$MCD [dB]}&
\multicolumn{2}{c|}{$\uparrow$LFC}&
\multicolumn{2}{c|}{$\downarrow$CER [\%]}&
\multicolumn{2}{c}{$\uparrow$pMOS}
\\\hline
\multicolumn{1}{c|}{s}&\multicolumn{1}{c V{3}}{t}&DPM&DPM+BNF&DPM&DPM+BNF&DPM&\!\!\!DPM+BNF\!\!\!&DPM&DPM+BNF\\\thline
      &   bdl&$8.25\pm .12$&$\bm{6.51\pm .09}$&$0.38\pm .05$&$\bm{0.52\pm .03}$&$13.7$&$\bm{2.4}$&$3.16\pm .05$&$\bm{3.44\pm .05}$\\
   clb&   slt&$6.34\pm .07$&$\bm{6.12\pm .06}$&$\bm{0.71\pm .03}$&$0.62\pm .03$&$7.2$&$\bm{2.3}$&$\bm{3.78\pm .06}$&$3.72\pm .05$\\
      &   rms&$7.49\pm .08$&$\bm{6.13\pm .06}$&$0.10\pm .05$&$\bm{0.56\pm .02}$&$12.8$&$\bm{3.2}$&$3.00\pm .05$&$\bm{3.59\pm .05}$\\\hline
      &   clb&$7.08\pm .11$&$\bm{5.89\pm .07}$&$0.17\pm .04$&$\bm{0.57\pm .03}$&$22.8$&$\bm{3.4}$&$3.04\pm .06$&$\bm{3.63\pm .05}$\\
   bdl&   slt&$7.18\pm .09$&$\bm{6.20\pm .05}$&$0.41\pm .04$&$\bm{0.63\pm .02}$&$16.9$&$\bm{3.0}$&$3.43\pm .05$&$\bm{3.64\pm .05}$\\
      &   rms&$7.33\pm .14$&$\bm{6.46\pm .12}$&$0.44\pm .04$&$\bm{0.48\pm .03}$&$15.6$&$\bm{3.5}$&$\bm{3.77\pm .05}$&$3.55\pm .05$\\\hline
      &   clb&$6.24\pm .08$&$\bm{5.79\pm .05}$&$\bm{0.69\pm .03}$&$0.59\pm .03$&$11.6$&$\bm{1.5}$&$\bm{3.70\pm .05}$&$3.65\pm .05$\\
   slt&   bdl&$7.93\pm .12$&$\bm{6.54\pm .08}$&$\bm{0.55\pm .04}$&$0.53\pm .03$&$10.1$&$\bm{2.2}$&$3.23\pm .05$&$\bm{3.46\pm .06}$\\
      &   rms&$7.80\pm .11$&$\bm{6.28\pm .09}$&$0.20\pm .05$&$\bm{0.53\pm .03}$&$12.1$&$\bm{2.0}$&$2.92\pm .05$&$\bm{3.60\pm .05}$\\\hline
      &   clb&$7.05\pm .08$&$\bm{6.02\pm .06}$&$0.07\pm .05$&$\bm{0.52\pm .03}$&$14.8$&$\bm{1.2}$&$3.10\pm .07$&$\bm{3.69\pm .05}$\\
   rms&   bdl&$7.95\pm .17$&$\bm{6.91\pm .14}$&$0.29\pm .05$&$\bm{0.44\pm .04}$&$9.8$&$\bm{1.1}$&$\bm{3.72\pm .05}$&$3.51\pm .05$\\
      &   slt&$7.51\pm .11$&$\bm{6.44\pm .09}$&$0.08\pm .05$&$\bm{0.55\pm .03}$&$13.7$&$\bm{1.2}$&$3.27\pm .05$&$\bm{3.64\pm .05}$\\\hline
\multicolumn{2}{c V{3}}{All pairs}
             &$7.35\pm .04$&$\bm{6.27\pm .03}$&$0.34\pm .02$&$\bm{0.55\pm .01}$&$13.4$&$\bm{2.2}$&$3.34\pm .02$&$\bm{3.59\pm .01}$\\\thline   
\end{tabular}
\end{scriptsize}
\end{table*}

\subsection{Objective Evaluation Metrics}


The test set for the above experiment
consisted of speech samples of each speaker reading the same sentences.
We evaluated the objective quality of the converted speech samples using mel-cepstral distortion (MCD) [dB], log $F_0$ correlation coefficient (LFC), and character error rate (CER) [\%].
We also evaluated the audio quality of the converted speech samples with a mean opinion score (MOS) predictor. 
We refer to this measure as the pseudo MOS (pMOS).

The utterance-level MCD was computed by averaging the frame-level MCDs along the dynamic time warping (DTW) path aligning the MCC vector sequences of the converted and target speech. 
LFC was also computed based on this DTW path. 
CER was evaluated using the wav2vec 2.0 model \cite{Baevski2020} (the ``Large LV-60K'' architecture with an extra linear module), pre-trained on 60,000 hours of unlabeled audio from Libri-Light \cite{Kahn2020} dataset, and fine-tuned on 960 hours of transcribed audio from LibriSpeech dataset \cite{Panayotov2015}.
To obtain the pMOS of the converted speech, Saeki's system \cite{Saeki2022} submitted to the VoiceMOS challenge 2022 \cite{Huang2022}, which exhibited a strong correlation with human MOS ratings, was used as the MOS predictor. 
The lower the value of MCD, the closer to 1 the LFC, the closer to 0 the CER, and the closer to 5 the pMOS, the better the performance.

\subsection{Comparison of DSM and DPM Formulations}

First, we compare the performance of the DSM and DPM versions of VoiceGrad.
\reftab{dsm_dpm_comp} shows the average utterance-level MCD and LFC, CER, and pMOS with 95\% confidence intervals of the converted speech obtained with the DSM and DPM versions of VoiceGrad. 
As the results show,  the DPM version performed slightly better than the DSM version for all the metrics.
Considering these results and the fact that the DPM version required only 11 iterations to perform the conversion, while the DSM version required 576 iterations, the DPM version was confirmed to be superior in our current implementation.
However, given that the CER for the ground truth target speech was only 1.1\%, we found that both versions tended to produce speech with relatively low intelligibility in terms of CER. 
In the following, we show that the idea of the BNF conditioning can significantly improve the CER of the generated speech.

\subsection{Effect of BNF Conditioning}

We evaluated the effect of the BNF conditioning described in \refsubsec{bnf_conditioning}.
\reftabs{bnf_conditioning_comp_dsm}{bnf_conditioning_comp_dpm} display the performance of the DSM and DPM versions of VoiceGrad with and without BNF conditioning, where `DSM+BNF' and `DPM+BNF' refer to the DSM and DPM versions with BNF conditioning, respevtively, while `s' and `t' denote source and target, respectively.
As the results show, the incorporation of the BNF sequence into the score approximator resulted in a significant performance improvement in terms of all of the metrics, especially in CER.
Significant improvements were also observed in terms of LFC, MCD, and pMOS, confirming that BNF conditioning contributes not only to intelligibility but also to the intonation, speaker similarity, and audio quality of the generated speech. 
This suggests that linguistic-related features can facilitate the prediction of the target score function.
Another finding was that while the version without BNF conditioning performed relatively less effectively in inter-gender conversions compared to intra-gender conversions, BNF conditioning improved the conversions, making them less gender-dependent.

\subsection{Comparison with Baseline Methods}
\label{subsec:baseline_comp}

\begin{table}[t!]
\caption{MCD [dB] Comparisons with Baseline Methods}
\label{tab:mcd_baseline_comp}
\vspace{-2ex}
\centering
\begin{scriptsize}
(a) Closed-set scenario
\begin{tabular}{l|l V{3} c|c|c|c}
\thline
\multicolumn{2}{c V{3}}{Speakers}&
\multirow{2}{*}{StarGAN-VC}&
\multirow{2}{*}{AutoVC}&
\multirow{2}{*}{PPG-VC}&
\multirow{2}{*}{VoiceGrad}
\\\cline{1-2}
\multicolumn{1}{c|}{s}&\multicolumn{1}{c V{3}}{t}&&&&\\\thline
      &   bdl&$8.10\pm .11$&$9.48\pm .09$&$7.86\pm .10$&$\bm{6.51\pm .09}$\\
   clb&   slt&$6.68\pm .07$&$8.63\pm .09$&$7.82\pm .09$&$\bm{6.12\pm .06}$\\
      &   rms&$7.55\pm .07$&$9.81\pm .08$&$7.75\pm .08$&$\bm{6.13\pm .06}$\\\hline
      &   clb&$7.73\pm .11$&$9.15\pm .15$&$7.76\pm .11$&$\bm{5.89\pm .07}$\\
   bdl&   slt&$7.85\pm .09$&$8.67\pm .08$&$8.00\pm .11$&$\bm{6.20\pm .05}$\\
      &   rms&$8.06\pm .13$&$8.64\pm .10$&$8.30\pm .09$&$\bm{6.46\pm .12}$\\\hline
      &   clb&$6.52\pm .06$&$8.67\pm .07$&$7.47\pm .08$&$\bm{5.79\pm .05}$\\
   slt&   bdl&$8.17\pm .09$&$9.17\pm .10$&$7.90\pm .10$&$\bm{6.54\pm .08}$\\
      &   rms&$8.14\pm .10$&$9.82\pm .08$&$8.19\pm .09$&$\bm{6.28\pm .09}$\\\hline
      &   clb&$7.82\pm .08$&$9.05\pm .16$&$7.62\pm .08$&$\bm{6.02\pm .06}$\\
   rms&   bdl&$8.93\pm .14$&$9.59\pm .09$&$8.41\pm .16$&$\bm{6.91\pm .14}$\\
      &   slt&$8.51\pm .11$&$8.94\pm .10$&$8.20\pm .13$&$\bm{6.44\pm .09}$\\\hline
\multicolumn{2}{c V{3}}{All pairs}
             &$7.84\pm .04$&$9.13\pm .04$&$7.94\pm .03$&$\bm{6.27\pm .03}$\\\thline   
\end{tabular}
\\\medskip
\centering
(b) Open-set scenario
\begin{tabular}{l|l V{3} c|c|c|c}
\thline
\multicolumn{2}{c V{3}}{Speakers}&
\multirow{2}{*}{StarGAN-VC}&
\multirow{2}{*}{AutoVC}&
\multirow{2}{*}{PPG-VC}&
\multirow{2}{*}{VoiceGrad}
\\\cline{1-2}
\multicolumn{1}{c|}{s}&\multicolumn{1}{c V{3}}{t}&&&&\\\thline
\multirow{4}{*}{jmk}
      &   clb&$8.44\pm .11$&$8.78\pm .16$&$7.86\pm .10$&$\bm{6.30\pm .08}$\\
      &   bdl&$8.48\pm .12$&$9.10\pm .14$&$7.86\pm .10$&$\bm{6.76\pm .10}$\\
      &   slt&$8.42\pm .08$&$8.33\pm .10$&$7.82\pm .09$&$\bm{6.40\pm .07}$\\
      &   rms&$7.66\pm .11$&$8.44\pm .11$&$7.75\pm .08$&$\bm{6.57\pm .10}$\\\hline
\multirow{4}{*}{ksp}
      &   clb&$8.70\pm .10$&$9.14\pm .16$&$7.76\pm .11$&$\bm{6.67\pm .09}$\\
      &   bdl&$9.64\pm .12$&$9.63\pm .09$&$7.76\pm .11$&$\bm{7.30\pm .12}$\\
      &   slt&$8.68\pm .11$&$8.93\pm .09$&$8.00\pm .11$&$\bm{6.81\pm .07}$\\
      &   rms&$8.18\pm .08$&$8.62\pm .10$&$8.30\pm .09$&$\bm{6.61\pm .07}$\\\hline
\multirow{4}{*}{lnh}
      &   clb&$7.18\pm .11$&$9.23\pm .18$&$7.47\pm .08$&$\bm{6.15\pm .09}$\\
      &   bdl&$8.16\pm .17$&$9.89\pm .11$&$7.90\pm .10$&$\bm{6.69\pm .13}$\\
      &   slt&$7.40\pm .10$&$8.93\pm .12$&$7.90\pm .10$&$\bm{6.10\pm .05}$\\
      &   rms&$8.08\pm .10$&$8.75\pm .13$&$8.19\pm .09$&$\bm{6.23\pm .07}$\\\hline
\multicolumn{2}{c V{3}}{All pairs}
             &$8.25\pm .04$&$8.98\pm .04$&$8.02\pm .04$&$\bm{6.55\pm .03}$\\\thline   
\end{tabular}
\end{scriptsize}
\end{table}

\begin{table}[t!]
\caption{LFC Comparisons with Baseline Methods}
\label{tab:lfc_baseline_comp}
\vspace{-2ex}
\centering
\begin{scriptsize}
(a) Closed-set scenario
\begin{tabular}{l|l V{3} c|c|c|c}
\thline
\multicolumn{2}{c V{3}}{Speakers}&
\multirow{2}{*}{StarGAN-VC}&
\multirow{2}{*}{AutoVC}&
\multirow{2}{*}{PPG-VC}&
\multirow{2}{*}{VoiceGrad}
\\\cline{1-2}
\multicolumn{1}{c|}{s}&\multicolumn{1}{c V{3}}{t}&&&&\\\thline
      &   bdl&$0.43\pm .05$&$0.23\pm .06$&$0.49\pm .04$&$\bm{0.52\pm .03}$\\
   clb&   slt&$0.71\pm .03$&$0.37\pm .06$&$\bm{0.64\pm .04}$&$0.62\pm .03$\\
      &   rms&$0.03\pm .04$&$0.25\pm .05$&$0.35\pm .05$&$\bm{0.56\pm .02}$\\\hline
      &   clb&$0.44\pm .05$&$0.02\pm .06$&$0.51\pm .04$&$\bm{0.57\pm .03}$\\
   bdl&   slt&$0.56\pm .04$&$0.17\pm .06$&$0.55\pm .05$&$\bm{0.63\pm .02}$\\
      &   rms&$0.12\pm .04$&$0.11\pm .05$&$0.28\pm .06$&$\bm{0.46\pm .03}$\\\hline
      &   clb&$\bm{0.69\pm .03}$&$0.26\pm .06$&$0.62\pm .04$&$0.59\pm .03$\\
   slt&   bdl&$0.43\pm .05$&$0.19\pm .07$&$\bm{0.53\pm .04}$&$\bm{0.53\pm .03}$\\
      &   rms&$0.09\pm .04$&$0.20\pm .06$&$0.24\pm .06$&$\bm{0.53\pm .03}$\\\hline
      &   clb&$0.41\pm .04$&$0.03\pm .06$&$\bm{0.57\pm .03}$&$0.52\pm .03$\\
   rms&   bdl&$0.18\pm .06$&$0.27\pm .05$&$0.43\pm .06$&$\bm{0.44\pm .04}$\\
      &   slt&$0.25\pm .05$&$0.46\pm .04$&$0.52\pm .05$&$\bm{0.55\pm .03}$\\\hline
\multicolumn{2}{c V{3}}{All pairs}
             &$0.36\pm .02$&$0.21\pm .02$&$0.48\pm .01$&$\bm{0.55\pm .01}$\\\thline   
\end{tabular}
\\\medskip
\centering
(b) Open-set scenario
\begin{tabular}{l|l V{3} c|c|c|c}
\thline
\multicolumn{2}{c V{3}}{Speakers}&
\multirow{2}{*}{StarGAN-VC}&
\multirow{2}{*}{AutoVC}&
\multirow{2}{*}{PPG-VC}&
\multirow{2}{*}{VoiceGrad}
\\\cline{1-2}
\multicolumn{1}{c|}{s}&\multicolumn{1}{c V{3}}{t}&&&&\\\thline
\multirow{4}{*}{jmk}
      &   clb&$0.25\pm .04$&$0.04\pm .06$&$\bm{0.56\pm .04}$&$\bm{0.56\pm .03}$\\
      &   bdl&$0.50\pm .05$&$0.10\pm .06$&$\bm{0.57\pm .04}$&$0.50\pm .03$\\
      &   slt&$0.33\pm .05$&$0.15\pm .05$&$\bm{0.62\pm .04}$&$0.61\pm .03$\\
      &   rms&$0.39\pm .05$&$0.05\pm .05$&$0.35\pm .05$&$\bm{0.53\pm .03}$\\\hline
\multirow{4}{*}{ksp}
      &   clb&$0.33\pm .05$&$-0.01\pm .06$&$0.45\pm .05$&$\bm{0.51\pm .03}$\\
      &   bdl&$0.08\pm .06$&$0.13\pm .06$&$0.34\pm .06$&$\bm{0.42\pm .03}$\\
      &   slt&$0.28\pm .05$&$0.28\pm .05$&$0.42\pm .05$&$\bm{0.53\pm .03}$\\
      &   rms&$0.17\pm .04$&$0.17\pm .05$&$0.26\pm .05$&$\bm{0.49\pm .03}$\\\hline
\multirow{4}{*}{lnh}
      &   clb&$\bm{0.59\pm .03}$&$-0.05\pm .05$&$0.54\pm .04$&$0.55\pm .03$\\
      &   bdl&$0.31\pm .04$&$0.08\pm .06$&$0.39\pm .05$&$\bm{0.49\pm .03}$\\
      &   slt&$0.59\pm .03$&$0.17\pm .05$&$0.57\pm .04$&$\bm{0.60\pm .03}$\\
      &   rms&$0.10\pm .03$&$0.07\pm .05$&$0.33\pm .04$&$\bm{0.52\pm .03}$\\\hline
\multicolumn{2}{c V{3}}{All pairs}
             &$0.33\pm .02$&$0.10\pm .02$&$0.45\pm .01$&$\bm{0.53\pm .01}$\\\thline   
\end{tabular}
\end{scriptsize}
\end{table}

\begin{table}[t!]
\caption{CER [\%] Comparisons with Baseline Methods}
\label{tab:cer_baseline_comp}
\vspace{-2ex}
\centering
\begin{scriptsize}
(a) Closed-set scenario
\begin{tabular}{l|l V{3} c|c|c|c}
\thline
\multicolumn{2}{c V{3}}{Speakers}&
\multirow{2}{*}{StarGAN-VC}&
\multirow{2}{*}{AutoVC}&
\multirow{2}{*}{PPG-VC}&
\multirow{2}{*}{VoiceGrad}
\\\cline{1-2}
\multicolumn{1}{c|}{s}&\multicolumn{1}{c V{3}}{t}&&&&\\\thline
      &   bdl&$3.77$&$74.96$&$3.44$&$\bm{2.44}$\\
   clb&   slt&$\bm{1.59}$&$73.89$&$3.03$&$2.35$\\
      &   rms&$7.26$&$71.57$&$4.10$&$\bm{3.17}$\\\hline
      &   clb&$5.25$&$71.56$&$3.96$&$\bm{3.40}$\\
   bdl&   slt&$5.69$&$72.06$&$4.08$&$\bm{2.96}$\\
      &   rms&$4.19$&$71.95$&$4.29$&$\bm{3.55}$\\\hline
      &   clb&$\bm{1.18}$&$71.70$&$2.19$&$1.47$\\
   slt&   bdl&$5.33$&$73.68$&$\bm{2.11}$&$2.21$\\
      &   rms&$14.55$&$70.47$&$2.60$&$\bm{1.97}$\\\hline
      &   clb&$4.95$&$71.89$&$\bm{1.15}$&$1.17$\\
   rms&   bdl&$1.32$&$74.68$&$1.34$&$\bm{1.09}$\\
      &   slt&$6.35$&$78.10$&$1.54$&$\bm{1.15}$\\\hline
\multicolumn{2}{c V{3}}{All pairs}
             &$5.12$&$73.04$&$2.82$&$\bm{2.24}$\\\thline   
\end{tabular}
\\\medskip
(b) Open-set scenario
\begin{tabular}{l|l V{3} c|c|c|c}
\thline
\multicolumn{2}{c V{3}}{Speakers}&
\multirow{2}{*}{StarGAN-VC}&
\multirow{2}{*}{AutoVC}&
\multirow{2}{*}{PPG-VC}&
\multirow{2}{*}{VoiceGrad}
\\\cline{1-2}
\multicolumn{1}{c|}{s}&\multicolumn{1}{c V{3}}{t}&&&&\\\thline
\multirow{4}{*}{jmk}
      &   clb&$14.48$&$72.71$&$\bm{3.01}$&$3.51$\\
      &   bdl&$3.53$&$73.31$&$\bm{2.71}$&$3.48$\\
      &   slt&$14.88$&$73.80$&$3.06$&$\bm{3.03}$\\
      &   rms&$\bm{3.26}$&$72.03$&$3.40$&$3.59$\\\hline
\multirow{4}{*}{ksp}
      &   clb&$25.18$&$72.05$&$\bm{10.64}$&$12.45$\\
      &   bdl&$\bm{10.78}$&$75.28$&$11.43$&$13.36$\\
      &   slt&$26.79$&$76.96$&$\bm{12.67}$&$13.93$\\
      &   rms&$13.44$&$71.35$&$13.11$&$\bm{12.01}$\\\hline
\multirow{4}{*}{lnh}
      &   clb&$2.13$&$74.19$&$\bm{2.11}$&$2.22$\\
      &   bdl&$5.45$&$76.00$&$\bm{2.35}$&$2.40$\\
      &   slt&$2.19$&$75.40$&$\bm{1.94}$&$2.16$\\
      &   rms&$10.16$&$74.24$&$\bm{2.54}$&$3.03$\\\hline
\multicolumn{2}{c V{3}}{All pairs}
             &$11.02$&$73.94$&$\bm{5.75}$&$6.26$\\\thline   
\end{tabular}
\end{scriptsize}
\label{tab:cer_baseline_comp_open}
\end{table}

\begin{table}[t!]
\caption{pMOS Comparisons with Baseline Methods}
\label{tab:pmos_baseline_comp}
\vspace{-2ex}
\centering
\begin{scriptsize}
(a) Closed-set scenario
\begin{tabular}{l|l V{3} c|c|c|c}
\thline
\multicolumn{2}{c V{3}}{Speakers}&
\multirow{2}{*}{StarGAN-VC}&
\multirow{2}{*}{AutoVC}&
\multirow{2}{*}{PPG-VC}&
\multirow{2}{*}{VoiceGrad}
\\\cline{1-2}
\multicolumn{1}{c|}{s}&\multicolumn{1}{c V{3}}{t}&&&&\\\thline
      &   bdl&$2.44\pm .05$&$1.24\pm .00$&$\bm{3.58\pm .05}$&$3.44\pm .05$\\
   clb&   slt&$3.67\pm .06$&$1.32\pm .01$&$3.70\pm .06$&$\bm{3.72\pm .06}$\\
      &   rms&$2.52\pm .05$&$1.30\pm .01$&$3.25\pm .06$&$\bm{3.59\pm .05}$\\\hline
      &   clb&$2.44\pm .06$&$1.26\pm .01$&$3.58\pm .05$&$\bm{3.63\pm .05}$\\
   bdl&   slt&$2.53\pm .06$&$1.34\pm .01$&$\bm{3.68\pm .06}$&$3.64\pm .05$\\
      &   rms&$2.85\pm .06$&$1.26\pm .01$&$3.38\pm .07$&$\bm{3.55\pm .05}$\\\hline
      &   clb&$3.75\pm .06$&$1.28\pm .01$&$3.54\pm .06$&$\bm{3.65\pm .05}$\\
   slt&   bdl&$2.31\pm .05$&$1.25\pm .01$&$\bm{3.52\pm .05}$&$3.46\pm .06$\\
      &   rms&$2.44\pm .06$&$1.31\pm .01$&$3.02\pm .07$&$\bm{3.60\pm .05}$\\\hline
      &   clb&$2.20\pm .06$&$1.25\pm .00$&$\bm{3.75\pm .05}$&$3.69\pm .06$\\
   rms&   bdl&$2.99\pm .04$&$1.23\pm .00$&$\bm{3.79\pm .06}$&$3.51\pm .05$\\
      &   slt&$2.27\pm .06$&$1.30\pm .01$&$\bm{3.73\pm .06}$&$3.64\pm .05$\\\hline
\multicolumn{2}{c V{3}}{All pairs}
             &$2.70\pm .03$&$1.28\pm .00$&$3.54\pm .02$&$\bm{3.59\pm .02}$\\\thline   
\end{tabular}
\\\medskip
(b) Open-set scenario
\begin{tabular}{l|l V{3} c|c|c|c}
\thline
\multicolumn{2}{c V{3}}{Speakers}&
\multirow{2}{*}{StarGAN-VC}&
\multirow{2}{*}{AutoVC}&
\multirow{2}{*}{PPG-VC}&
\multirow{2}{*}{VoiceGrad}
\\\cline{1-2}
\multicolumn{1}{c|}{s}&\multicolumn{1}{c V{3}}{t}&&&&\\\thline
\multirow{4}{*}{jmk}
      &   clb&$2.10\pm .05$&$1.26\pm .01$&$3.70\pm .05$&$\bm{3.71\pm .05}$\\
      &   bdl&$3.23\pm .05$&$1.25\pm .01$&$\bm{3.78\pm .05}$&$3.55\pm .05$\\
      &   slt&$2.16\pm .06$&$1.40\pm .02$&$\bm{3.79\pm .06}$&$3.66\pm .06$\\
      &   rms&$3.37\pm .05$&$1.26\pm .01$&$3.50\pm .05$&$\bm{3.69\pm .04}$\\\hline
\multirow{4}{*}{ksp}
      &   clb&$2.32\pm .06$&$1.26\pm .01$&$3.45\pm .05$&$\bm{3.59\pm .06}$\\
      &   bdl&$2.82\pm .06$&$1.24\pm .00$&$\bm{3.56\pm .05}$&$3.36\pm .05$\\
      &   slt&$2.46\pm .05$&$1.37\pm .01$&$3.48\pm .06$&$\bm{3.50\pm .05}$\\
      &   rms&$2.43\pm .05$&$1.26\pm .01$&$3.13\pm .06$&$\bm{3.55\pm .05}$\\\hline
\multirow{4}{*}{lnh}
      &   clb&$2.92\pm .06$&$1.28\pm .01$&$3.70\pm .05$&$\bm{3.77\pm .06}$\\
      &   bdl&$2.48\pm .04$&$1.27\pm .01$&$\bm{3.73\pm .05}$&$3.56\pm .05$\\
      &   slt&$2.77\pm .06$&$1.41\pm .02$&$3.72\pm .06$&$\bm{3.76\pm .05}$\\
      &   rms&$2.44\pm .05$&$1.28\pm .01$&$3.36\pm .06$&$\bm{3.65\pm .05}$\\\hline
\multicolumn{2}{c V{3}}{All pairs}
             &$2.62\pm .02$&$1.30\pm .00$&$3.57\pm .02$&$\bm{3.61\pm .02}$\\\thline   
\end{tabular}
\end{scriptsize}
\end{table}

\begin{table*}[t!]
\caption{MCD [dB], LFC, CER [\%], and pMOS of Source Speech}
\label{tab:source_speech}
\vspace{-2ex}
\centering
\begin{scriptsize}
\begin{tabular}{l|l V{3} c|c|c|c}
\thline
\multicolumn{2}{c V{3}}{Speakers}&
\multirow{2}{*}{$\downarrow$MCD [dB]}&
\multirow{2}{*}{$\uparrow$LFC}&
\multirow{2}{*}{$\downarrow$CER [\%]}&
\multirow{2}{*}{$\uparrow$pMOS}
\\\cline{1-2}
\multicolumn{1}{c|}{s}&\multicolumn{1}{c V{3}}{t}& & & & \\\thline
      &   bdl&$9.62\pm .11$&$0.30\pm .06$&\multirow{3}{*}{$1.22$}&\multirow{3}{*}{$4.34$}\\
   clb&   slt&$7.52\pm .07$&$0.63\pm .04$&&\\
      &   rms&$9.95\pm .09$&$0.30\pm .05$&&\\\hline
      &   clb&$9.62\pm .11$&$0.30\pm .06$&\multirow{3}{*}{$1.67$}&\multirow{3}{*}{$4.24$}\\
   bdl&   slt&$9.69\pm .10$&$0.35\pm .06$&&\\
      &   rms&$9.58\pm .12$&$0.32\pm .06$&&\\\hline
      &   clb&$7.52\pm .07$&$0.63\pm .04$&\multirow{3}{*}{$0.76$}&\multirow{3}{*}{$4.36$}\\
   slt&   bdl&$9.69\pm .10$&$0.35\pm .06$&&\\
      &   rms&$9.91\pm .08$&$0.30\pm .06$&&\\\hline
      &   clb&$9.95\pm .09$&$0.30\pm .05$&\multirow{3}{*}{$0.71$}&\multirow{3}{*}{$4.40$}\\
   rms&   bdl&$9.58\pm .12$&$0.32\pm .06$&&\\
      &   slt&$9.91\pm .08$&$0.30\pm .06$&&\\\hline
\multirow{4}{*}{jmk}
      &   clb&$9.66\pm .09$&$0.11\pm .07$&\multirow{4}{*}{$1.47$}&\multirow{4}{*}{$4.06$}\\
      &   bdl&$9.61\pm .10$&$0.46\pm .06$&&\\
      &   slt&$9.57\pm .08$&$0.25\pm .07$&&\\
      &   rms&$8.95\pm .10$&$0.37\pm .06$&&\\\hline
\multirow{4}{*}{ksp}
      &   clb&$9.80\pm .10$&$0.16\pm .07$&\multirow{4}{*}{$2.73$}&\multirow{4}{*}{$3.99$}\\
      &   bdl&$10.22\pm .12$&$0.10\pm .07$&&\\
      &   slt&$9.53\pm .10$&$0.21\pm .06$&&\\
      &   rms&$8.72\pm .08$&$0.28\pm .05$&&\\\hline
\multirow{4}{*}{lnh}
      &   clb&$8.68\pm .12$&$0.56\pm .04$&\multirow{4}{*}{$1.28$}&\multirow{4}{*}{$4.24$}\\
      &   bdl&$8.50\pm .17$&$0.33\pm .06$&&\\
      &   slt&$8.87\pm .12$&$0.47\pm .06$&&\\
      &   rms&$9.13\pm .09$&$0.30\pm .05$&&\\\thline
\end{tabular}
\end{scriptsize}
\end{table*}

\begin{table*}[t!]
\caption{MCD [dB], LFC, CER [\%], and pMOS Obtained with Diff-VC \cite{Popov2022}}
\label{tab:diff-vc}
\vspace{-2ex}
\centering
\begin{scriptsize}
(a) Closed-set scenario\\
\begin{tabular}{l|l V{3} c|c|c|c}
\thline
\multicolumn{2}{c V{3}}{Speakers}&
\multirow{2}{*}{$\downarrow$MCD [dB]}&
\multirow{2}{*}{$\uparrow$LFC}&
\multirow{2}{*}{$\downarrow$CER [\%]}&
\multirow{2}{*}{$\uparrow$pMOS}
\\\cline{1-2}
\multicolumn{1}{c|}{s}&\multicolumn{1}{c V{3}}{t}& & & & \\\thline
      &   bdl&$7.75\pm .10$&$0.26\pm .05$&$2.81$&$3.86\pm .05$\\
   clb&   slt&$7.44\pm .08$&$0.30\pm .05$&$2.90$&$3.97\pm .05$\\
      &   rms&$7.75\pm .10$&$0.21\pm .04$&$3.39$&$3.77\pm .05$\\\hline
      &   clb&$9.62\pm .11$&$0.30\pm .06$&$3.99$&$3.76\pm .06$\\
   bdl&   slt&$8.45\pm .16$&$0.22\pm .05$&$3.61$&$3.86\pm .06$\\
      &   rms&$8.83\pm .15$&$0.15\pm .05$&$4.02$&$3.67\pm .05$\\\hline
      &   clb&$8.03\pm .12$&$0.25\pm .04$&$2.70$&$3.85\pm .06$\\
   slt&   bdl&$7.73\pm .10$&$0.28\pm .05$&$2.30$&$3.85\pm .05$\\
      &   rms&$8.18\pm .13$&$0.19\pm .04$&$3.06$&$3.71\pm .04$\\\hline
      &   clb&$8.08\pm .10$&$0.26\pm .05$&$2.41$&$3.85\pm .05$\\
   rms&   bdl&$8.41\pm .15$&$0.20\pm .06$&$2.24$&$3.84\pm .05$\\
      &   slt&$7.79\pm .13$&$0.27\pm .05$&$2.58$&$3.96\pm .05$\\\thline
\multicolumn{2}{c V{3}}{All pairs}
             &$8.01\pm .04$&$0.24\pm .01$&$3.00$&$3.83\pm .02$\\\thline 
\end{tabular}
\\\medskip
\centering
(b) Open-set scenario\\
\begin{tabular}{l|l V{3} c|c|c|c}
\thline
\multicolumn{2}{c V{3}}{Speakers}&
\multirow{2}{*}{$\downarrow$MCD [dB]}&
\multirow{2}{*}{$\uparrow$LFC}&
\multirow{2}{*}{$\downarrow$CER [\%]}&
\multirow{2}{*}{$\uparrow$pMOS}
\\\cline{1-2}
\multicolumn{1}{c|}{s}&\multicolumn{1}{c V{3}}{t}& & & & \\\thline
\multirow{4}{*}{jmk}
      &   clb&$8.11\pm .11$&$0.24\pm .05$&$4.66$&$3.96\pm .03$\\
      &   bdl&$8.03\pm .13$&$0.24\pm .06$&$4.35$&$3.97\pm .05$\\
      &   slt&$7.55\pm .09$&$0.31\pm .05$&$4.08$&$4.06\pm .04$\\
      &   rms&$8.16\pm .11$&$0.18\pm .06$&$4.65$&$3.87\pm .04$\\\hline
\multirow{4}{*}{ksp}
      &   clb&$8.51\pm .10$&$0.22\pm .05$&$32.70$&$3.75\pm .05$\\
      &   bdl&$8.89\pm .15$&$0.19\pm .06$&$31.83$&$3.73\pm .05$\\
      &   slt&$8.26\pm .11$&$0.25\pm .05$&$33.19$&$3.97\pm .06$\\
      &   rms&$8.16\pm .08$&$0.13\pm .05$&$28.93$&$3.60\pm .05$\\\hline
\multirow{4}{*}{lnh}
      &   clb&$7.94\pm .12$&$0.31\pm .04$&$3.09$&$3.99\pm .05$\\
      &   bdl&$7.84\pm .16$&$0.24\pm .05$&$3.25$&$3.97\pm .05$\\
      &   slt&$7.19\pm .09$&$0.34\pm .05$&$2.52$&$4.03\pm .05$\\
      &   rms&$7.74\pm .11$&$0.24\pm .04$&$2.99$&$3.80\pm .04$\\\thline
\multicolumn{2}{c V{3}}{All pairs}
             &$8.03\pm .04$&$0.24\pm .01$&$13.02$&$3.88\pm .01$\\\thline 
\end{tabular}
\end{scriptsize}
\end{table*}

\reftabss{mcd_baseline_comp}{pmos_baseline_comp} show the MCD, LFC, CER, and pMOS results of the proposed and baseline methods under the closed-set and open-set conditions. Note that the proposed method here refers to the DPM version with BNF conditioning.
For reference, the MCD, LFC, CER, and pMOS of each source speaker's speech are shown in \reftab{source_speech}.

As the results indicate, among the baseline methods, PPG-VC excelled in nearly all metrics, particularly in CER and pMOS, highlighting its capability to produce highly intelligible and high-quality speech. Remarkably, VoiceGrad slightly surpassed this top-performing baseline in terms of CER and pMOS and notably outperformed it in MCD and LFC.
This suggests that the speech produced by VoiceGrad is as clear and natural as or even more so than that produced by PPG-VC, while having features more similar to the target speaker.
This was also confirmed by our subjective evaluation test, which will be discussed in detail in the following section.
StarGAN-VC performed one step below PPG-VC and VoiceGrad in overall performance. Nevertheless, the low CER confirms that StarGAN-VC's intelligibility in the generated speech was comparable to PPG-VC and VoiceGrad.
AutoVC performed worse than the other methods, even with the authors' implementation as-is, likely due to its strong dependence on dataset domains for optimal architecture and hyperparameters. While some improvement might have been possible with specific tuning for the CMU ARCTIC dataset, surpassing PPG-VC appears challenging for AutoVC, as PPG-VC has consistently shown to outperform AutoVC in both audio quality and speaker similarity according to the results in VCC2020.

While all the tested methods are expected to handle any-to-many conversions, it is crucial to evaluate their robustness to speech input from unknown speakers. 
This can be confirmed in the differences in each method's performance between the closed-set and open-set conditions.
Among the tested methods, AutoVC and PPG-VC exhibited little difference in performance between closed-set and open-set conditions. In contrast, StarGAN-VC and VoiceGrad showed slight performance degradation in MCD and CER in the open-set condition. This indicates the high robustness of AutoVC and PPG-VC in converting speech input from unknown speakers but suggests issues with StarGAN-VC and VoiceGrad in this regard. Nevertheless, VoiceGrad's performance was the best among the tested methods, except for CER.
In terms of CER, the converted speech from the source speaker ksp consistently exhibited a relatively high error rate for all the methods. This can be attributed to the fact that speaker ksp is an Indian-accented English speaker and indicates challenges with handling accented speech in all the methods.

As a reference, we also evaluated the MCD, LFC, CER, and pMOS for the speech converted by Diff-VC \cite{Popov2022} under the same conditions as above. The results are shown in \reftab{diff-vc}. According to the results, while Diff-VC showed impressive performance in terms of pMOS, VoiceGrad showed superior performance in all the other metrics.
Despite VoiceGrad being built on seemingly strong assumptions, as pointed out by the authors of Diff-VC, its success hinges on a crucial factor: how closely the distribution of the log mel-spectrogram difference between source and target utterances resembles a Gaussian. 
While its validity remains uncertain, 
the fact that VoiceGrad performs reasonably well may suggest that this assumption is not entirely inaccurate.

\subsection{Real-Time Factor for Mel-Spectrogram Conversion}

The real-time factor of the computation time required for the mel-spectrogram conversion for each of the tested methods is shown in \reftab{rtf_baseline_comp}.
As \reftab{rtf_baseline_comp} shows, 
the DSM version of VoiceGrad was considerably slower than the other methods, while the DPM version was nearly as fast as PPG-VC. This was due to the cosine-based noise variance scheduling, which allowed the reverse diffusion process with considerably fewer steps to produce high-quality mel-spectrograms.
We also found that BNF conditioning did not have a significant impact on computation time.
The fastest was StarGAN-VC, followed by AutoVC.
All algorithms were implemented in PyTorch and run on a single Tesla V100 SXM2 GPU with a 32.0 GB memory and an Intel(R) Xeon(R) Gold 5218 16-core CPU @ 2.30GHz.

\begin{table}[t!]
\caption{Real-Time Factor Comparisons}
\label{tab:rtf_baseline_comp}
\begin{tabular}{c|c|c|c|c|c}
\thline
\multirow{2}{*}{\!\!StarGAN-VC\!\!}&\multirow{2}{*}{AutoVC}&\multirow{2}{*}{PPG-VC}&\multicolumn{3}{c}{VoiceGrad}\\\cline{4-6}
&&&DSM&DPM&\!\!DPM+BNF\!\!\\\hline
0.0014&0.0057&0.0245&1.180&0.0221&0.0235\\\thline
\end{tabular}
\end{table}

\subsection{Subjective Listening Tests}

\begin{table*}[t!]
\caption{Results of the MOS Test}
\vspace{-2ex}
\centering
\begin{tabular}{l V{3} c | c | c | c | c | c}
\thline
\multirow{2}{*}{Scenario}&
\multirow{2}{*}{Real}&
\multirow{2}{*}{StarGAN-VC}&
\multirow{2}{*}{AutoVC}&
\multirow{2}{*}{PPG-VC}&
\multicolumn{2}{c}{VoiceGrad}
\\\cline{6-7}
& & & & &DPM&DPM+BNF
\\\hline
Closed-set&
\multirow{2}{*}{$4.82\pm .12$}&
$2.66\pm .09$&
$1.19\pm .05$&
$3.65\pm .11$&
$3.22\pm .10$&
$\bm{3.77\pm .11}$
\\
Open-set&
&
$2.61\pm .08$&
$1.23\pm .05$&
$3.70\pm .10$&
$3.13\pm .10$&
$\bm{3.81\pm .10}$
\\\thline
\end{tabular}
\label{tab:MOS_qlt}
\end{table*}

\begin{table*}[t!]
\caption{Results of Speaker Similarity Test}
\vspace{-2ex}
\centering
\begin{tabular}{l V{3} c | c | c | c | c }
\thline
\multirow{2}{*}{Scenario}&
\multirow{2}{*}{StarGAN-VC}&
\multirow{2}{*}{AutoVC}&
\multirow{2}{*}{PPG-VC}&
\multicolumn{2}{c}{VoiceGrad}
\\\cline{5-6}
& & & &DPM&DPM+BNF
\\\hline
Closed-set&
$2.56\pm .10$&
$1.87\pm .09$&
$1.88\pm .10$&
$2.35\pm .14$&
$\bm{3.58\pm .07}$
\\
Open-set&
$2.02\pm .09$&
$1.82\pm .09$&
$1.92\pm .11$&
$2.67\pm .10$&
$\bm{3.67\pm .06}$
\\\thline
\end{tabular}
\label{tab:MOS_sim}
\end{table*}

We conducted mean opinion score (MOS) tests to compare the audio quality 
and speaker similarity 
of the converted speech samples 
generated by the proposed and baseline methods.
For the proposed method, we included samples of both the DPM version with and without BNF conditioning. 
For these tests, we used samples obtained under the close-set and open-set conditions, as in \refsubsec{baseline_comp}.
Sixteen listeners 
participated in both tests.
The tests were conducted online, where each participant was asked 
to use a headphone in a quiet environment.

With the audio quality test, 
we included the real speech samples of the target speakers as reference. 
Each listener was  
asked to evaluate the naturalness of each sample on a five-point scale 
by selecting 5: Excellent, 4: Good, 3: Fair, 2: Poor, or 1: Bad for each utterance.
The scores with 95\% confidence intervals
are shown in \reftab{MOS_qlt}.
With the speaker similarity test, 
each listener was given a converted speech sample and 
a real speech sample of the corresponding target speaker 
and asked to evaluate how likely they were produced by the same speaker on a four-point scale by selecting 
4: Same (sure), 3: Same (not sure), 2: Different (not sure), or 1: Different (sure). 
The scores with 95\% confidence intervals
are shown in \reftab{MOS_sim}.

Comparing the versions with and without BNF conditioning in VoiceGrad, the former was significantly higher in both audio quality and speaker similarity, confirming the effectiveness of BNF conditioning.
Upon listening to the actual speech samples, it becomes clear that there is often noticeable phoneme distortion in the speech converted without BNF conditioning. In contrast, the version with BNF conditioning demonstrates a significant improvement in intelligibility.
Among the baseline methods, PPG-VC performed best in terms of audio quality but fell short of StarGAN-VC in terms of speaker similarity, and AutoVC performed worst in both tests. 
The BNF-conditioned version of VoiceGrad performed better than PPG-VC in audio quality and better than StarGAN-VC in speaker similarity. These were consistent with the results of the quantitative evaluation described in the previous section.
The version without BNF conditioning performed reasonably well, with better audio quality than StarGAN-VC, though not as good as PPG-VC, and better speaker similarity than PPG-VC, though slightly less than StarGAN-VC. However, it was only in the closed-set condition that the version without BNF conditioning was not as good as StarGAN-VC in terms of speaker similarity, and in the open-set condition it performed better than StarGAN-VC.
Comparing the results of both tests under closed-set and open-set conditions, all the methods performed similarly in the closed and open set conditions, with the exception of StarGAN-VC in the speaker similarity test.

It is noteworthy that while BNF conditioning demonstrated only a modest effect of 0.2 to 0.4 in the quantitative evaluation of pMOS, it yielded a more substantial impact of 0.5 to 0.7 in the subjective MOS evaluation. This divergence in results is anticipated, as the MOS predictor likely emphasizes the assessment of audio quality (naturalness as a speech waveform) over intelligibility (the accuracy of phoneme pronunciation). Conversely, participants in the subjective listening test likely took into account both audio quality and intelligibility when evaluating the naturalness of the stimuli.

In summary, these results indicate from subjective listening tests that VoiceGrad (1) outperforms the tested baselines in both audio quality and speaker similarity, (2) is greatly improved by BNF conditioning, and (3) works as well for unknown speaker input as for known speaker input.

Audio examples of these methods
are provided at \cite{Kameoka2023url_voicegrad}.

\section{Conclusion}

In this paper, we proposed VoiceGrad, 
a non-parallel any-to-many VC method
based upon the concepts of score matching, Langevin dynamics, and DPMs:
The idea involves training a score approximator, a fully convolutional network with a U-Net structure, to predict the gradient of the log density of the mel-spectrograms of multiple speakers. Once the network is trained, it can be used to perform VC through annealed Langevin dynamics or reverse diffusion process to iteratively update the mel-spectrogram of input speech to sound like a target speaker's voice.
Through objective and subjective experiments, VoiceGrad has demonstrated superior performance to the tested baselines in terms of audio quality and speaker similarity under both closed-set and open-set conditions. Additionally, we have found that the concept of BNF conditioning significantly enhances the intelligibility of the generated speech.


\appendix

\section{Appendix}

$\nabla_{\Vec{x}_l}\log q(\Vec{x}_l) = \mathbb{E}_{\Vec{x}_0}[\nabla_{\Vec{x}_l} \log q(\Vec{x}_l|\Vec{x}_0)]$
can be proved as
\begin{align}
\nabla_{\Vec{x}_l}\log q(\Vec{x}_l)
&=
\frac{
\nabla_{\Vec{x}_l}
q(\Vec{x}_l)
}{
q(\Vec{x}_l)
}
\nonumber\\
&=
\frac{
\nabla_{\Vec{x}_l}
\int 
q(\Vec{x}_l|\Vec{x}_0)q(\Vec{x}_0) d\Vec{x}_0
}{
q(\Vec{x}_l)
}
\nonumber\\
&=
\frac{
\int
q(\Vec{x}_0) 
\nabla_{\Vec{x}_l}
q(\Vec{x}_l|\Vec{x}_0)
d\Vec{x}_0
}{
q(\Vec{x}_l)
}
\nonumber\\
&=
\frac{
\int
q(\Vec{x}_0) 
\nabla_{\Vec{x}_l} \exp (\log q(\Vec{x}_l|\Vec{x}_0))
d\Vec{x}_0
}{
q(\Vec{x}_l)
}
\nonumber\\
&=
\frac{
\int
q(\Vec{x}_0) 
q(\Vec{x}_l|\Vec{x}_0)
\nabla_{\Vec{x}_l}
\log q(\Vec{x}_l|\Vec{x}_0)
d\Vec{x}_0
}{
q(\Vec{x}_l)
}
\nonumber\\
&=
\int
q(\Vec{x}_0|\Vec{x}_l)
\nabla_{\Vec{x}_l}
\log q(\Vec{x}_l|\Vec{x}_0)
d\Vec{x}_0
\nonumber\\
&=
\mathbb{E}_{\Vec{x}_0\sim q(\Vec{x}_0|\Vec{x}_l)}[
\nabla_{\Vec{x}_l}
\log q(\Vec{x}_l|\Vec{x}_0)
],
\end{align}
where the third line follows from the Leibniz integral rule, the fifth line follows from the chain rule, and the sixth line follows from the Bayes rule.



\ifCLASSOPTIONcaptionsoff
  \newpage
\fi



\bibliographystyle{IEEEtran}
\bibliography{Kameoka2023IEEETrans_VoiceGrad}
%

\end{document}